\documentclass[rmp,aps]{revtex4}
\usepackage{graphicx}
\usepackage{float}
\usepackage{mathptmx} 
\usepackage{natbib}
\usepackage[section]{placeins}

\begin{document}

\title{
Fitting and goodness-of-fit test of non-truncated and truncated
power-law distributions
}
\author
{
\'Alvaro Corral$^{\dag}$\email{ACorral at crm.cat}
and Anna Deluca$^{\dag *}$\email{ADeluca at crm.cat}  
}
\affiliation{
$^\dag$%
Centre de Recerca Matem\`atica,
Edifici C, Campus Bellaterra,
E-08193, Barcelona, Spain\\
$^*$%
Departament de Matem\`atiques, 
Universitat Aut\`onoma de Barcelona,
E-08193 Cerdanyola, Spain 
}
\date{\today}

\maketitle

\section*{Abstract}

Power-law distributions contain precious information
about a large variety of processes in geoscience and elsewhere.
Although there are sound theoretical grounds for these distributions, 
the empirical evidence in favor of power laws
has been traditionally weak.
Recently, Clauset et al. have proposed a systematic method 
to find over which range (if any) a certain distribution behaves as
a power law.
However, their method has been found to fail,
in the sense that true (simulated) power-law tails are not recognized as such
in some instances, 
and then the power-law hypothesis is rejected.
Moreover, the method does not work well 
when extended to power-law distributions 
with an upper truncation.
We explain in detail a similar but alternative procedure,
valid for truncated as well as for non-truncated power-law distributions,
based in maximum likelihood estimation, the Kolmogorov-Smirnov
goodness-of-fit test, and Monte Carlo simulations.
An overview of the main concepts
as well as a recipe for their practical implementation is provided.
The performance of our method is put to test
on several empirical data which were previously analyzed 
with less systematic approaches.
The databases presented here include the half-lives of 
the radionuclides, the seismic moment of earthquakes in the whole 
world and in Southern California, a proxy for the energy dissipated
by tropical cyclones elsewhere, the area burned by forest fires in Italy,
and the waiting times calculated over different spatial subdivisions 
of Southern California.
We find the functioning of the method very satisfactory.


\section{Introduction}

Over the last decades, the importance of power-law distributions 
has continuously increased, not only in geoscience but elsewhere
\citep{Johnson_univariate_continuous}.
These are probability distributions defined by a probability density 
(for a continuous variable $x$) or by a probability mass function
(for a discrete variable $x$) given by,
\begin{equation}
f(x) \propto \frac 1 {x^\alpha},
\end{equation}
for $x \ge a$ and $a>0$,
with a normalization factor (hidden in the proportionality symbol $\propto$)
which depends on whether $x$ is continuous or discrete.
In any case, normalization implies $\alpha > 1$.
Sometimes power-law distributions are also called Pareto distributions
\cite{Johnson_univariate_continuous,Evans}
(or Riemann zeta distributions in the discrete case 
\cite{Johnson_univariate}),
although in other contexts the name Pareto is associated to a
slightly different distribution \cite{Johnson_univariate_continuous}.
So we stick to the clearer term power-law distribution.

These have remarkable, non-usual statistical properties, 
as are scale invariance and divergence of moments.
The first one means that power-law functions (defined between 0 and $\infty$) 
are invariant under (properly performed) linear rescaling of axes (both $x$ and $f$)
and therefore have no characteristic scale, and hence cannot be used to define
a prototype of the observable represented by $x$ 
\cite{Takayasu,Christensen_Moloney,Newman_05,Corral_Lacidogna}. 
For example, no unit of distance 
can be defined from the gravitational field of a point mass (a power law),
whereas a time unit can be defined for radioactive decay (an exponential function).
However, as
power-law distributions cannot be defined for all $x>0$ but for $x \ge  a > 0$
their scale invariance is not ``complete'' or strict.

A second uncommon property is the non-existence of finite moments;
for instance, if $\alpha \le 2$ not a single finite moment exists (no mean, no variance, etc.).
This has important consequences, as the law of large numbers does not hold
\cite[p. 65]{Kolmogorov}, 
i.e., the mean of a sample does not converge {to a finite value} 
as the size of the sample increases; {rather, the sample mean tends to infinite}
\cite[p. 393]{Shiryaev}.
If $2 < \alpha \le 3$ the mean exists and is finite, but higher moments
are infinite, which means for instance that the central limit theorem,
in its classic formulation, does not apply 
(the mean of a sample is not normally distributed
and has infinite standard deviation)
\cite{Bouchaud_Georges}.
Higher $\alpha$'s yield higher-order moments infinite, but
then the situation is not so ``critical''.
Newman reviews other peculiar properties of power-law distributions,
such as the 80$|$20 rule 
\cite{Newman_05}.

Although the normal (or Gaussian) distribution gives a non-zero probability that a 
human being is 10 m or 10 km tall, 
the definition of the probability density up to infinity is not {questionable} at all,
and the same happens with an exponential distribution and most ``standard'' distributions
in probability theory.
However, one already sees that the power-law distribution {is} problematic,
in particular for $\alpha \le 2$, as it predicts an infinite mean,
and for $2 \le \alpha < 3$, as the variability of the sample mean is infinite.
Of course, there can be variables having an infinite mean
(one can easily simulate in a computer processes in which the time between events
has an infinite mean), but in other cases, for physical reasons,
the mean should be finite.
In such situations a simple generalization is the truncation of the tail 
\cite{Johnson_univariate_continuous,Burroughs_Tebbens,Aban}, 
yielding the truncated power-law distribution, defined in the same way as before by
$f(x) \propto 1/x^\alpha$ but with $a \le x \le  b$, with $b$ finite, and with normalizing factor
depending now on $a$ and $b$ (in some cases it is possible to have $a=0$, see next section).
Obviously, the existence of a finite upper cutoff $b$ 
automatically leads to well-behaved moments,
if the statistics is enough to ``see'' the cutoff; 
on the other hand, a range of scale invariance can
persist, if $b \gg a$.
What one finds in some practical problems is that the statistics is not enough to decide
which is the sample mean and one cannot easily conclude if a pure power law or a truncated power
law is the right model for the data.

A well known example of (truncated or not) 
power-law distribution is the Gutenberg-Richter law for earthquake ``size''
\cite{Utsu_GR,Kagan_gji02,Kanamori_rpp}.
If by size we understand radiated energy, the Gutenberg-Richter law  implies
that, in any seismically active region of the world,
the sizes of earthquakes follow a power-law distribution, 
with an exponent $\alpha=1+2B/3$ and $B$ close to 1.
In this case, scale invariance means that 
{if one asks}
{how big (in terms of radiated energy) earthquakes are in a certain region,}
{such a simple question} has no possible answer.
The non-convergence of the mean energy can easily be checked
from data: catastrophic events such as the Sumatra-Andaman mega-earthquake of 2004
contribute to the mean much more than the previous recorded history
\cite{Corral_FontClos}.
Note that for the most common formulation of the Gutenberg-Richter law,
in terms of the magnitude,
earthquakes are not power-law distributed, but this is due
to the fact that magnitude is an (increasing) exponential function of radiated
energy, and therefore magnitude turns out to be exponentially distributed.
In terms of magnitude, the statistical properties of earthquakes are trivial
(well behaved mean, existence of a characteristic magnitude...),
but we insist that this is not the case in terms of radiated energy.

\citet{Malamud_hazards} lists several other natural hazards following power-law distributions
in some (physical) measure of size,
such as 
rockfalls, 
landslides \cite{Hergarten_book}, volcanic eruptions \cite{McClelland,Lahaie}, and forest fires \cite{Malamud_fires_pnas},
and we can add rainfall \cite{Peters_prl,Peters_Deluca}, tropical cyclones (roughly speaking, hurricanes) \cite{Corral_hurricanes},
auroras \cite{Freeman_Watkins}, tsunamis \cite{Burroughs}, etc.
In some cases this broad range of responses is triggered simply by a small driving
or perturbation
(the slow motion of tectonic plates for earthquakes, the continuous pumping of solar radiation
in hurricanes, etc.); then, 
this highly nonlinear relation between input and output
can be labeled as crackling noise \cite{Sethna_nature}.
Notice that this does not apply for tsunamis, for instance, as they are not slowly driven
(or at least not directly slowly driven).

\citet{Aschwanden_open} reviews
disparate astrophysical phenomena which are distributed according
to power laws, some of them related to geoscience: sizes of asteroids, craters in the Moon, solar flares, 
and energy of cosmic rays.  
In the field of ecology and close areas,
the applicability of power-law distributions
has been overviewed by \citet{White},
mentioning also island and lake sizes.
\citet{Aban} provides bibliography
for power-law and other heavy-tailed distributions
in diverse disciplines, including hydrology,
and \citet{Burroughs_Tebbens} provide interesting geological examples.

A theoretical framework for power-law distributed sizes (and durations) of catastrophic phenomena
not only in geoscience but also in condensed matter physics, astrophysics, biological evolution,
neuroscience,
and even the economy, is provided by the concept of self-organized criticality, 
and summarized by the sandpile paradigm
\cite{Bak_book,Jensen,Sornette_critical_book,Christensen_Moloney,Pruessner_book}.
However, although the ideas of self-organization and criticality
are very reasonable in the context of most of the geosystems mentioned above
\cite{Peters_np,Peters_jh,Corral_Elsner},
one cannot rule out other mechanisms for the emergence of power-law distributions
\cite{Sornette_critical_book,Mitz,Newman_05,Dickman_rain,Czechowski}.

On the other hand, it is interesting to mention that, in addition to
sizes and durations, power-law distributions have also been extensively 
reported in time between the occurrences of natural hazards
(waiting times), as for instance in
solar flares \cite{Boffetta,Baiesi_flares},
earthquakes \cite{Bak.2002,Corral_pre.2003,Corral_physA.2004},
or solar wind \cite{Wanliss};
in other cases the distributions contain a power-law part
mixed with other factors
\cite{Corral_prl.2004,Saichev_Sornette_times,Geist_Parsons,Corral_test}.
Nevertheless, the possible relation with critical phenomena is not direct
\cite{Paczuski_btw,Corral_comment}.
The distance between events, or jumps, has received relatively 
less attention \cite{Davidsen_distance,Corral_prl.2006,Felzer}.

The importance of power-law distributions in geoscience is apparent;
however, some of the evidence gathered in favor of this paradigm
can be considered as ``anecdotic'' or tentative, as it is based on rather poor data
analysis. A common practice is to find some (naive or not) estimation
of the probability density or mass function $f(x)$ and plot $\ln f(x)$ versus $\ln x$
and look for a linear dependence between both variables.
Obviously, a power-law distribution should behave in that way, 
but the opposite is not true: an apparent straight line in a log-log plot
of $f(x)$ should not be considered a guarantee of an underlying power-law distribution,
or perhaps the exponent obtained from there is clearly biased
\cite{Goldstein,Bauke,White,Clauset}.
But in order to discriminate between several competing theories or models, 
as well as in order to extrapolate the available statistics to the most
extreme events, it is very important to 
properly fit power laws
and to find the right power-law exponent (if any) \cite{White}.

The subject of this paper is a discussion on the most appropriate
fitting, testing of the goodness-of-fit, and representation of power-law distributions, 
both non-truncated
and truncated. A consistent and robust method will be checked on several examples
in geoscience, including earthquakes, tropical cyclones, and forest fires. 
The procedure is in some points analogous to that of \citet{Clauset},
although there are variations is some key steps, in order to correct several 
drawbacks of the original method \cite{Corral_nuclear,Peters_Deluca}. 
The most important difference is in the criterion 
to select the range over which the power law holds.
As the case of most interest in geoscience is that of a continuous random variable, 
the more involving discrete case will be postponed to a separate publication 
\cite{Corral_Deluca_arxiv,Corral_Boleda}.

\section{Power-law fits and goodness-of-fit tests}

\subsection{Non-truncated and truncated power-law distributions}

Let us consider a continuous power-law distribution, defined in the range
$a \le x \le b$, where  
$b$ can be finite or infinite
and
$a \ge 0$.
The probability density of $x$ is given by,
\begin{equation}
f(x)=\frac{\alpha-1}{a^{1-\alpha}-1/b^{\alpha-1}}\left(\frac 1 x\right)^\alpha,
\end{equation}
the limit 
$b\rightarrow \infty$
provides the non-truncated power-law distribution
if $\alpha>1$ and $a>0$,
also called here pure power law;
otherwise, for finite $b$ 
one has the truncated power law,
for which no restriction exists on $\alpha$
if $a>0$,
but $\alpha < 1$ if $a=0$ 
(which is sometimes referred to as the power-function distribution \cite{Evans});
the case $\alpha=1$ needs a separate treatment, with
\begin{equation}
f(x)=\frac 1 {x\ln (b/a)}.
\end{equation}
We will consider in principle that the distribution has a unique parameter, $\alpha$,
and that $a$ and $b$ are fixed and known values.
Remember that, at point $x$, the probability density function of a random variable 
is defined as the probability per unit of the variable that 
the random variable lies in a infinitesimal interval around $x$,
that is,
\begin{equation}
f(x) = \lim_{\Delta x \rightarrow 0}
\frac{\mbox{Prob}[x \le \mbox{random variable} < x+\Delta x]}{\Delta x},
\end{equation}
and has to verify $f(x)\ge 0$ and $\int_{-\infty}^{\infty} f(x) dx=1$, 
see for instance \citet{Ross_firstcourse}.

Equivalently, we could have characterized the distribution 
by its (complementary) cumulative distribution function,
\begin{equation}
S(x) = \mbox{Prob}[\mbox{random variable} \ge  x] = \int_x^\infty f(x') dx'.
\end{equation}
For a truncated or non-truncated power law this leads to
\begin{equation}
S(x)=
\frac{1/x^{\alpha-1}-1/b^{\alpha-1}}{a^{1-\alpha}-1/b^{\alpha-1}},
\end{equation}
if $\alpha \ne 1$ and
\begin{equation}
S(x)=\frac{\ln(b/x)}{\ln(b/a)},
\end{equation}
if $\alpha=1$.
Note that although $f(x)$ always has a power-law shape, 
$S(x)$ only has it in the non-truncated case ($b\rightarrow \infty$ and $\alpha>1$);
nevertheless, even not being a power law in the truncated case,
the distribution is a power law, as it is $f(x)$ and not $S(x)$
which gives the name to the distribution.

\subsection{Problematic fitting methods}

Given a set of data, there are many methods to 
fit a probability  distribution.
\citet{Goldstein},
\citet{Bauke},
\citet{White},
and 
\citet{Clauset} check several methods
based in the fitting of the estimated probability
densities or cumulative distributions
in the power-law case.
As mentioned in the first section, 
$\ln f(x)$ is then a linear function of $\ln x$,
both for non-truncated and truncated power laws.
The same holds for $\ln S(x)$, but only in the non-truncated case.
So, one can either estimate $f(x)$ from data, 
using some binning procedure, 
or estimate $S(x)$, for which no binning is necessary,
and then fit a straight line by the least-squares method.
As we find White et al.'s (2008) study the most complete,
we summarize their results below, 
although those of the other authors
are not very different.

For non-truncated power-law distributions,
White et al. (2008) find that the results of the least-squares method 
using the cumulative distribution are reasonable, 
although the points in $S(x)$ are not independent
and linear regression should yield problems in this case.
We stress that this procedure only can work for non-truncated
distributions (i.e., with $b\rightarrow \infty$),
truncated ones yield bad results \cite{Burroughs_Tebbens}.

The least-squares method applied to the probability density $f(x)$
has several variations, depending on the way of estimating $f(x)$.
Using linear binning one obtains a simple histogram, for which the
fitting results are catastrophic \cite{Goldstein,Bauke,Pueyo, White}. This is not unexpected, 
as linear binning of a heavy-tailed distribution can be considered
as a very naive approach.
If instead of linear binning one uses logarithmic binning
the results improve (when done ``correctly''), and are reasonable in some cases, 
but they still show some bias, high variance, and depend on the size of the bins.
A fundamental point should be to avoid having empty bins, 
as they are disregarded in logscale, 
introducing an important bias.

In summary, methods of estimation of probability-distribution parameters
based on least-squares fitting can have many problems, and usually
the results are biased. Moreover, these methods do not take into account
that the quantity to be fitted is a probability distribution
(i.e., once the distributions are estimated, 
the method is the same as for any other kind of function).
We are going to see that the method of maximum likelihood
is precisely designed for dealing with probability distributions,
presenting considerable advantages in front of
the other methods just mentioned.

\subsection{Maximum likelihood estimation}

Let us denote a sample of the random variable $x$ with $N$ elements
as $x_1$, $x_2$, \dots , $x_N$, and let us consider a probability
distribution $f(x)$ parameterized by $\alpha$. 
The likelihood function $L(\alpha)$ is defined as
the joint probability density (or the joint probability mass function
if the variable were discrete) evaluated at $x_1$, $x_2$, \dots , $x_N$
in the case in which the variables were independent, 
i.e., 
\begin{equation}
L(\alpha)=\prod_{i=1}^{N} f(x_i).
\end{equation}
Note that the sample is considered fixed, 
and it is the parameter $\alpha$ what is allowed to vary.
In practice it is more convenient to work with the log-likelihood, 
the natural logarithm of the likelihood (dividing by $N$ also, in our definition),
\begin{equation}
\ell(\alpha)=\frac 1 N \ln L(\alpha)=\frac 1 N \sum_{i=1}^{N} \ln f(x_i).
\end{equation}
The maximum likelihood (ML) estimator of the parameter $\alpha$ based on the sample
is just the maximum of $\ell(\alpha)$ as a function of $\alpha$ 
(which coincides with the maximum of $L(\alpha)$, obviously).
For a given sample, we will denote the ML estimator as $\alpha_e$
($e$ is from empirical), but it is important to realize
that the ML estimator is indeed a statistic
(a quantity calculated from a random sample) and therefore
can be considered as a random variable;
in this case it is denoted as $\hat \alpha$.
In a formula, 
\begin{equation}
\alpha_e=\mbox{argmax}_{\forall \alpha} \ell(\alpha),
\end{equation}
where argmax refers to the argument of the function $\ell$
that make it maximum.

For the truncated or the non-truncated continuous power-law distribution 
we have, substituting $f(x)$ from Eqs. (2)-(3) and introducing $r=a/b$,
disregarding the case $a=0$,
\begin{equation}
\ell(\alpha)= 
\ln \frac{\alpha-1}{1-r^{\alpha-1}} - \alpha \ln \frac g a - \ln a, \mbox{ if } \alpha \ne 1,
\end{equation}
\begin{equation}
\ell(\alpha)=
-\ln\ln \frac 1 r -\ln g, \mbox{ if } \alpha = 1;
\end{equation}
$g$ is the geometric mean of the data, $\ln g = N^{-1}\sum_1^N \ln x_i$,
and the last term in each expression is irrelevant for the maximization of $\ell(\alpha)$.
{The equation for $\alpha=1$ is necessary in order to avoid overflows in the
numerical implementation.}
Remember that the distribution is only parameterized by $\alpha$, whereas 
$a$ and $b$ (and $r$) are constant parameters;
therefore, $\ell(\alpha)$ is not a function of $a$ and $b$, but of $\alpha$.

In order to find the maximum of $\ell(\alpha)$ can derive with respect $\alpha$ and 
set the result equal to zero \cite{Johnson_univariate_continuous,Aban},
\begin{equation}
\left.\frac{d\ell(\alpha)}{d\alpha} \right|_{\alpha=\alpha_e}= \frac 1 {\alpha_e-1}+
\frac{r^{\alpha_e-1}\ln r}{1-r^{\alpha_e-1}}-\ln \frac g a=0,
\end{equation}
which constitutes the so-called likelihood equation for this problem.
For a non-truncated distribution, $r=0$,
and it is clear that there is one and only one solution,
\begin{equation}
\alpha_e=1+\frac 1 {\ln (g/a)},
\end{equation}
which corresponds to a maximum, 
as
\begin{equation}
L(\alpha)=e^{N\ell(\alpha)}=\frac 1 {a^N}(\alpha-1)^N e^{-N\alpha\ln(g/a)},
\end{equation}
has indeed a maximum (resembling a gamma probability density, see next subsection).
Figure 1 illustrates the log-likelihood function and its derivate, 
for simulated power-law data.

\begin{figure}[h]
\includegraphics*[width=13cm,angle=270]
{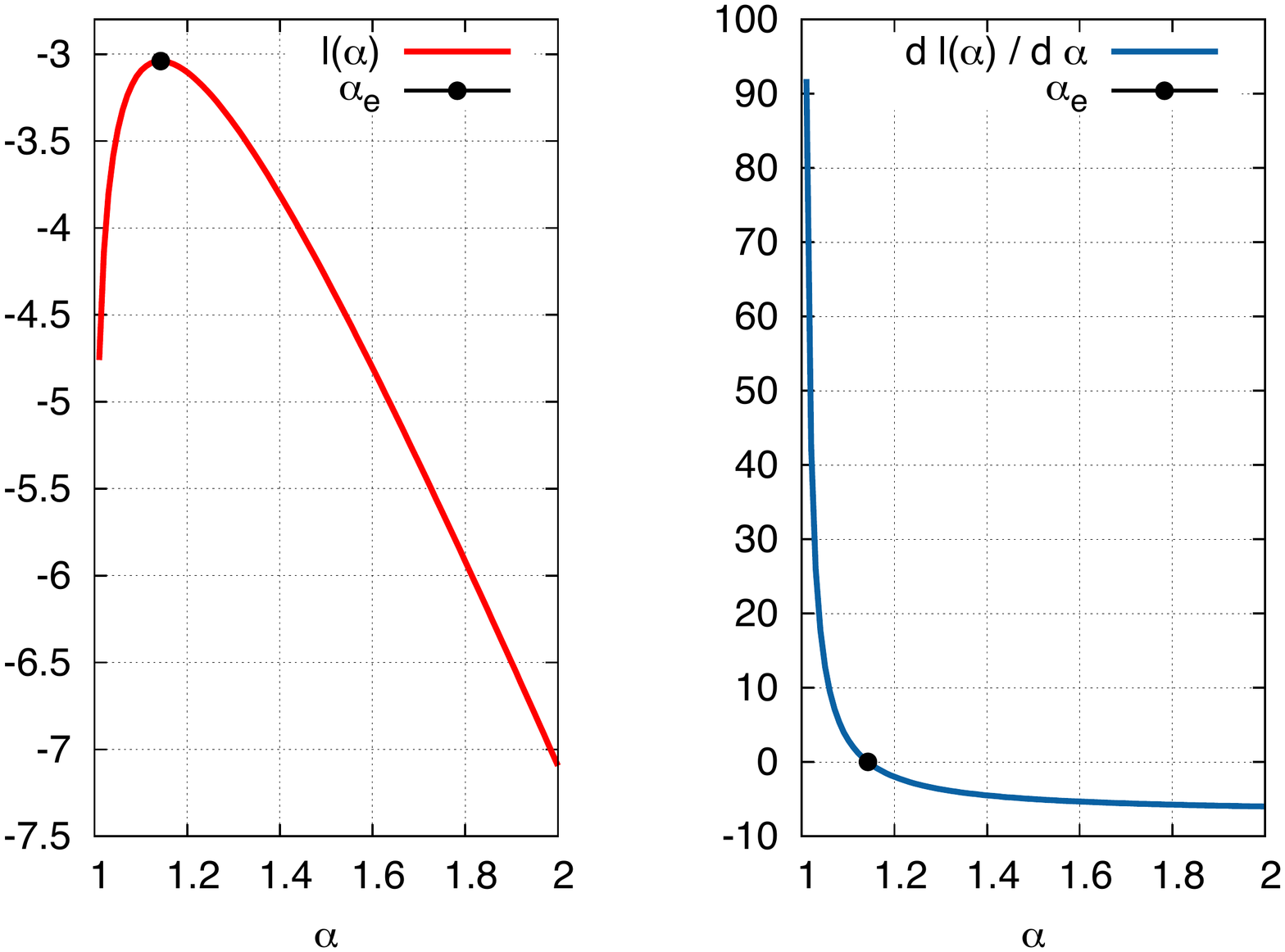}
\caption{Log-likelihood $\ell(\alpha)$ and its derivative, 
for simulated non-truncated power-law data with exponent $\alpha=1.15$
and $a=0.001$. The total number of data is $N_{tot}=1000$.
The resulting estimation leads to $\alpha_e=1.143$, 
which will lead to a confidence interval $\alpha \pm \sigma=1.143 \pm 0.005$.}  
\end{figure}

In the truncated case it is not obvious that there is
a solution to the likelihood equation \cite{Aban};
however, one can take advantage of the fact 
that the power-law distribution,
for fixed $a$ and $b$,
can be viewed as belonging to the regular exponential family,
for which it is known that the maximum likelihood
estimator exists and is unique, see \citet[p. 151]{Nielsen}, \citet{Castillo_unpublished}.
Indeed, in the single-parameter case,
the exponential family can be written in the form,
\begin{equation}
f(x)=C^{-1}(\alpha)H(x) e^{\theta(\alpha) \cdot T(x)},
\end{equation}
where both $\theta(\alpha)$ and $T(x)$ can be vectors,
the former containing the parameter $\alpha$ of the family.
Then, for $\theta(\alpha)=-\alpha$, $T(x)=\ln x$, and $H(x)=1$
we obtain the (truncated or not) power-law distribution,
which therefore belongs to the regular exponential family, 
which guarantees the existence of a unique ML solution.

%


In order to find the ML estimator of the exponent in the truncated case, we proceed 
by maximizing directly the log-likelihood $\ell(\alpha)$
(rather than by solving the likelihood equation). 
{The reason is a practical one, as our procedure is part of a more general method, 
valid for arbitrary distributions $f(x)$, for which the derivative of
$\ell(\alpha)$ can be difficult to evaluate.}
We will use the downhill simplex method,
{through the routine {\tt amoeba} of} \citet{Press}, although
any other simpler {maximization} procedure should work, as the problem is one-dimensional,
in this case. 
One needs to take care when the value of $\alpha$ gets very close to one
in the maximization algorithm, and then replace $\ell(\alpha)$ by its limit at $\alpha=1$,
\begin{equation}
\ell(\alpha)\rightarrow_{\alpha \rightarrow 1} 
-\ln\ln \frac 1 r  - \alpha \ln \frac g a - \ln a,
\end{equation}
which is in agreement with the likelihood function 
for a (truncated) power-law distribution with 
$\alpha=1$.

An important property of ML estimators, not present 
in other fitting methods, is their invariance 
under re-parameterization.
If instead of working with parameter $\alpha$
we use $\nu=h(\alpha)$, then, the ML estimator
of $\nu$ is ``in agreement'' with that of $\alpha$,
i.e., $\hat \nu=h(\hat \alpha)$.
Indeed,
\begin{equation}
\frac{d\ell}{d\nu}=\frac{d\ell}{d\alpha}\frac{d\alpha}{d\nu},
\end{equation}
so, the maximum of $\ell$ as a function of $\nu$ is attained
at the point $h(\hat\alpha)$, provided that the function
$h$ is ``one-to-one''.
Note that the parameters could be multidimensional as well.
\citet{Casella} studies this invariance with much more care.

In their comparative study,
White et al. (2008) conclude that maximum likelihood estimation outperforms
the other methods, as always yields the lowest variance and bias of the estimator.
This is not unexpected, as the ML estimator is, mathematically,
the one with minimum variance among all asymptotically unbiased estimators.
This property is called asymptotical efficiency \cite{White,Bauke}.


\subsection{Standard deviation of the ML estimator}

The main result of this subsection is the value
of the uncertainty $\sigma$ of $\hat \alpha$, 
{represented by the standard deviation of $\hat \alpha$ and}
given by \cite{Aban}
\begin{equation}
\sigma
=\frac 1 {\sqrt{N }}\left[\frac 1 {(\alpha_e-1)^2} - \frac{r^{\alpha_e-1}\ln^2 r}
{(1-r^{\alpha_e-1})^2}\right]^{-1/2}.
\end{equation}
{This formula can be used directly, although $\sigma$ can be computed 
as well from Monte Carlo simulations, as explained in another subsection.
A third option is the use of the jackknife procedure, 
as done by \cite{Peters_Deluca}.
The three methods lead to essentially the same results.}
The rest of this subsection is devoted to the particular derivation of 
$\sigma$ for a non-truncated power-law distribution,
and therefore can be skipped by readers {interested mainly in the practical
use of ML estimation}.

For the calculation of $\hat \alpha$ (the ML estimator of $\alpha$)
one needs to realize that this is indeed a statistic
(a quantity calculated from a random sample) and therefore
it can be considered as a random variable.
Note that $\alpha$ denotes the true value of the parameter,
which is unknown.
It is more convenient to work with $\alpha-1$ (the exponent
of the cumulative distribution function);
in the non-truncated case ($r=0$ with $\alpha > 1$) we can easily derive its distribution.
First let us consider the geometric mean of the sample, $g$,
rescaled by the minimum value $a$,
\begin{equation}
\ln \frac g a = \frac 1 N \sum_{i=1}^{N} \ln \frac{x_i} a.
\end{equation}
As each $x_i$ is power-law distributed (by hypothesis),
a simple change of variables shows that
$\ln (x_i/a)$ turns out to be exponentially distributed,
with scale parameter $1/(\alpha-1)$;
then, the sum will be gamma distributed with the same scale parameter 
and with shape parameter given by $N$ 
(this is the key property of the gamma distribution \cite{Durrett}).
Therefore, $\ln (g/a)$ will follow the same gamma distribution 
but with scale parameter $N^{-1}(\alpha-1)^{-1}$.

At this point it is useful to introduce the generalized gamma distribution 
\cite{Johnson_univariate_continuous,Evans,Kalbfleisch2},
with density, for a random variable $y \ge 0$,
\begin{equation}
D(y)=\frac {|\delta|}{c \Gamma(\gamma/\delta)}
\left(\frac{y}{c}\right)^{\gamma-1}
e^{-(y/c)^\delta},
\end{equation}
where $c>0$ is the scale parameter and $\gamma$ and $\delta$ are the
shape parameters, which have to verify $0 < \gamma/\delta < \infty$
(so, the only restriction is that they have the same sign,
although the previous references
only consider $\gamma >0$ and $\delta >0$);
the case $\delta=1$ yields the usual gamma distribution
and $\delta=\gamma=1$ is the exponential one.
Again, changing variables one can show that the inverse $z=1/y$
of a generalized gamma variable is also a generalized gamma variable, 
but with transformed parameters,
\begin{equation}
\gamma, \delta, c \rightarrow -\gamma, -\delta, \frac 1 c.
\end{equation}

So, $\hat \alpha-1=z=1/\ln(g/a)$ will have a generalized gamma
distribution, with parameters $-N$, $-1$, and $N(\alpha-1)$
(keeping the same order as above).
Introducing the moments of this distribution \cite{Evans},
\begin{equation}
\langle y^m\rangle 
=c^m\frac {\Gamma\left(\frac{\gamma+m}{\delta}\right)}
{\Gamma\left({\gamma}/{\delta}\right)}
\end{equation}
(valid for $m > -\gamma$ if $\gamma>0$
and for $m < |\gamma|$ if $\gamma <0$,
and $\langle y^m\rangle$ infinite otherwise),
we obtain the expected value of $\hat\alpha-1$,
\begin{equation}
\langle \hat \alpha-1 \rangle = \frac {N(\alpha-1)}{N-1}.
\end{equation}
Note that the ML estimator, $\hat \alpha$, is biased, as its expected value
does not coincide with the right value, $\alpha$; however, 
asymptotically, the right value is recovered.
An unbiased estimator 
of $\alpha$ can be obtained for a small sample 
as $(1-1/N) \alpha_e +1/N$,
although this will not be of interest for us.

In the same way, the standard deviation of $\hat \alpha -1$
(and of $\hat \alpha$) turns out to be
\begin{equation}
\sigma = 
\sqrt{\langle (\hat \alpha -1)^2 \rangle - \langle \hat \alpha -1 \rangle^2}=
\frac {\alpha-1}{(1-1/N)\sqrt{N-2}}, 
\end{equation}
which leads asymptotically to $(\alpha-1)/\sqrt{N}$. 
In practice, we need to replace $\alpha$ by the estimated 
value $\alpha_e$; then, this is nothing else 
than the limit $r=0$ ($b\rightarrow \infty$)
of the general formula stated above for $\sigma$
\cite{Aban}.
The fact that the standard deviation tends to zero 
asymptotically (together with the fact that 
the estimator is asymptotically unbiased) 
implies that any single estimation converges (in probability) to the 
true value, and therefore the estimator
is said to be consistent.


\subsection{Goodness-of-fit test}

One can realize that the maximum likelihood method
always yields a ML estimator for $\alpha$,
no matter which data one is using.
In the case of power laws, 
as the data only enters in the likelihood function
through its geometric mean, any sample with
a given geometric mean yields the same value
for the estimation, although the sample can come
from a true power law or from any other distribution.
So, no quality of the fit is guaranteed and thus, maximum likelihood estimation should be rather
called minimum unlikelihood estimation.
For this reason 
a goodness-of-fit test is necessary
(although recent works do not take into account this fact
\cite{White,Kagan_gji02,Baro_Vives}).

Following \citet{Goldstein} and \citet{Clauset} 
we use the Kolmogorov-Smirnov (KS) test
\cite{Press,Chicheportiche},
based on the calculation of the KS statistic or KS distance $d_e$
between the theoretical probability distribution, represented by $S(x)$,
and the empirical one, $S_e(x)$.
The latter, which is an unbiased estimator of the cumulative distribution
\cite{Chicheportiche},
is given by 
the stepwise function 
\begin{equation}
S_e(x)=n_e(x)/N,
\end{equation}
where $n_e(x)$ is the number of data in the sample taking a value
of the variable larger than or equal to
$x$.
The KS statistic is just the maximum difference, in absolute value, 
between $S(x)$ and $n_e(x)/N$, that is, 
\begin{equation}
d_e=\max_{a \le x \le b}\left|S(x) - S_e(x)\right|=
\max_{a \le x \le b}\left|
\frac 1 {1-r^{\alpha_e-1}} \left[ \left(\frac a x\right)^{\alpha_e-1}-r^{\alpha_e-1}\right]
 - \frac{n_e(x)}{N}\right|,
\end{equation}
where 
the bars denote absolute value.
Note that the theoretical cumulative distribution $S(x)$ is parameterized
by the value of $\alpha$ obtained from ML, $\alpha_e$.
In practice, the difference only needs to be evaluated {around} the points
$x_i$ of the sample
(as the routine {\tt ksone} of \citet{Press} does)
{and not for all $x$}.
A more strict mathematical definition uses the supremum 
instead of the maximum, but in practice the maximum
works perfectly.
We illustrate the procedure in Fig. 2, 
with a simulation of a non-truncated power law.

\begin{figure}[h]
\includegraphics*[height=10cm]
{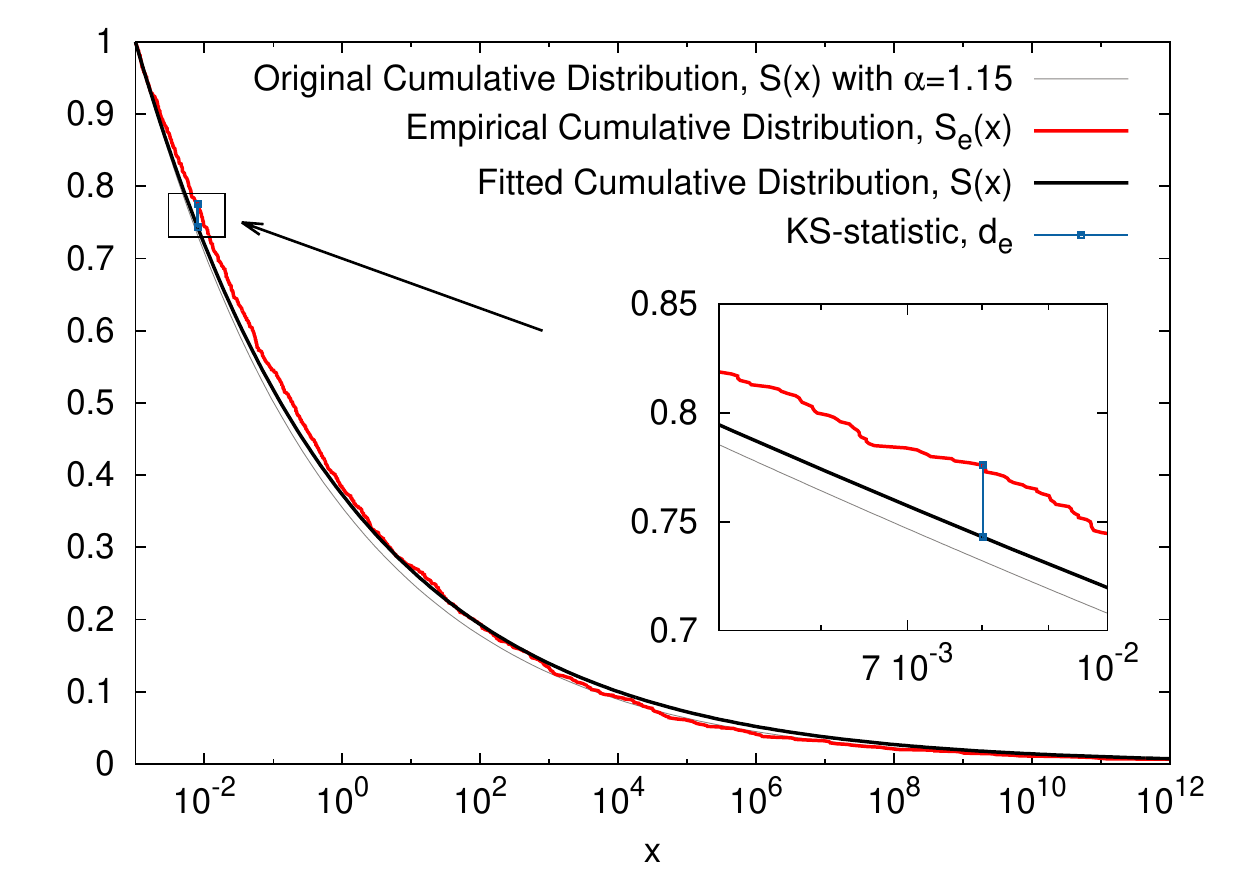}
\caption{Empirical (complementary) cumulative distribution 
for a simulated non-truncated power-law distribution with $\alpha=1.15$, $a=0.001$,
and $N_{tot}=1000$,
together with its corresponding fit, which yields $\alpha_e=1.143$.
The maximum difference between both curves, $d_e=0.033$, is marked
as an illustration of the calculation of the KS statistic.
The original theoretical distribution, unknown in practice, is also plotted.
}  
\end{figure}

Intuitively, if $d_e$ is large the fit is bad, 
whereas if $d_e$ is small the fit can be considered as good.
But the relative scale of $d_e$ is provided by its own 
probability distribution, through the calculation
of a $p-$value. 
Under the hypothesis that
the data follow indeed the theoretical distribution, 
with the parameter $\alpha$ obtained from our previous estimation
(this is the null hypothesis),
the $p-$value provides the probability that the
KS statistic takes a value larger than the one
obtained empirically,
i.e.,
\begin{equation}
p=\mbox{Prob}[\mbox{KS statistic for power-law data (with $\alpha_e$) is} > d_e];
\end{equation}
then, bad fits will have rather small $p-$values.

It turns out that, in principle, the distribution of the KS statistic 
is known, at least asymptotically, independently of the underlying form of the
distribution, so,
\begin{equation}
p_Q=Q(d_e\sqrt{N} + 0.12 d_e +0.11 d_e/\sqrt{N} )
=2\sum_{j=1}^\infty (-1)^{j-1} \exp[-2j^2 (d_e\sqrt{N} + 0.12 d_e +0.11 d_e/\sqrt{N} )^2],
\end{equation}
for which {one can} use the routine {\tt probks} of \citet{Press}
(but note that Eq. (14.3.9) there is not right). 
Nevertheless, this formula will not be accurate in our case,
{and for this reason we use the symbol $p_Q$ instead of $p$.
The reason is that}
we are ``optimizing'' the value of $\alpha$
using the same sample to which we apply the KS test, 
which yields a bias in the test, 
i.e., the formula would work for the true value of $\alpha$,
but not for one obtained by ML, which would yield in general a smaller
$KS$ statistic and too large $p-$values
(because the fit for $\alpha_e$ is better than for
the true value $\alpha$) \cite{Goldstein,Clauset}.
However, for this very same reason the formula can be
useful to reject the goodness of a given fit, 
i.e., if $p$ obtained in this way is already below 0.05,
the true $p$ will be even smaller and the fit is certainly bad.
But the opposite is not true.
In a formula,
\begin{equation}
\mbox{ if } p_Q < 0.05 \Rightarrow \mbox{ reject power law},
\end{equation}
otherwise, no decision can be taken yet.
{Of course, the significance level 0.05 is arbitrary and
can be changed to another value, as usual in statistical tests.}
As a final comment, perhaps a more
powerful test would be to use, instead of the KS statistic, 
the Kuiper's statistic \cite{Press}, which is a refinement of the former one. 
It is stated by Clauset et al. (2009) that both tests lead to very similar fits.
In most cases, we have also found no significant differences between both cases.


\subsection{The Clauset et al.'s recipe}

Now we are in condition to explain the genuine Clauset et al.'s (2009) method. 
{This is done in this subsection for completeness, and for the information of the reader, 
as we are not going to apply this method.}
The key to fitting a power law is neither the ML estimation of the 
exponent nor the goodness-of-fit test, but the selection of
the interval $[a,b]$ over which the power law holds.
Initially, we have taken $a$ and $b$ as fixed parameters, 
but in practice this is not the case, 
and one has to decide where the power law starts and where
ends, independently of the total range of the data.
In any case, $N$ will be the number of data in the power-law range
(and not the total number of data).

The recipe of Clauset et al. (2009) applies to non-truncated power-law
distributions ($b\rightarrow \infty$), and 
considers that $a$ is a variable which needs to be fit
from the sample
(values of $x$ below $a$ are outside the
power-law range). 
The recipe simply consists in 
the search of the value of $a$ which yields a 
minimum of the KS statistic, using as a parameter of the theoretical distribution
the one obtained by maximum likelihood, $\alpha_e$,
for the corresponding $a$
(no calculation of a $p-$value is required for each fixed $a$).
In other words,
\begin{equation}
a=\mbox{the one that yields minimum } d_e.
\end{equation}
Next, a global $p-$value is computed by 
generating synthetic samples by
a mixture of parametric bootstrap
{(similarly to what is explained in the next subsection)}
and
non-parametric bootstrap.
Then, the same procedure applied to the empirical data
(minimization of the KS distance using ML for fitting)
is applied to the syntetic samples
in order to fit $a$ and $\alpha$.

These authors do not provide any explanation of why
this should work, although one can argue that, if the data is indeed
a power law with the desired exponent, the larger the number of data
(the smaller the $a-$value), the smaller the value of $d_e$,
as $d_e$ goes as $1/\sqrt{N}$ (for large $N$, see previous subsection).
On the other hand, if for a smaller $a$ the data departs from the power law,
this deviation should compensate and overcome the reduction in 
$d_e$ due to the increase of $N$, yielding a larger $d_e$.
But there is no reason to justify this overcoming.

Nevertheless, we will not use the Clauset's et al.'s (2009) procedure for two other reasons.
First, its extension to truncated power laws, although obvious,
and justifiable with the same arguments, 
yields bad results, as the resulting values of the upper truncation cutoff, $b$, are
highly unstable.
Second, even for non-truncated distributions, it has been shown
that the method fails to detect the existence of a power law
for data simulated with a power-law tail \cite{Corral_nuclear}:
the method
yields an $a-$value well below the true power-law region, 
and therefore the resulting $p$ is too small 
for the power law to become acceptable.
We will explain an alternative method that avoids these problems,
but first let us come back to the case with $a$ and $b$ fixed.

\subsection{Monte Carlo simulations}

Remember that we are considering a power-law distribution,
defined in $a\le x \le b$.
We already have fit the distribution, by ML,
and we are testing the goodness of the fit
by means of the KS statistic. 
In order to obtain a reliable $p-$value for this test we will
perform Monte Carlo simulations of the whole process.
A synthetic sample power-law distributed and with $N$ elements
can be obtained in a straightforward way,
from the inversion 
or transformation 
method \cite{Press,Ross_firstcourse,Devroye},
\begin{equation}
x_i=\frac a {[1-(1-r^{\alpha_e-1})u_i]^{1/(\alpha_e-1)}},
\end{equation} 
where $u_i$ represents a uniform random number in $[0,1)$.
{One can use any random number generator for it.
Our results arise from {\tt ran3} of \citet{Press}.
}

\subsection{Application of the complete procedure to many syntetic samples
and calculation of $p-$value}

The previous fitting and testing procedure is applied in exactly the same way
to the synthetic sample, yielding a ML exponent $\alpha_s$
(where the subindex $s$ stands from synthetic or simulated), and then a KS statistic $d_s$,
computed as the difference between the theoretical cumulative
distribution, with parameter $\alpha_s$, and
the simulated one, $n_s(x)/N$ 
(obtained from simulations with $\alpha_e$, as described in the previous
subsection), i.e.,
\begin{equation}
d_s=\max_{a \le x \le b}\left|
\frac 1 {1-r^{\alpha_s-1}} \left[ \left(\frac a x\right)^{\alpha_s-1}-r^{\alpha_s-1}\right]
 - \frac{n_s(x)}{N}\right|.
\end{equation}
Both values of the exponent, $\alpha_e$ and $\alpha_s$, 
should be close {to each other}, but {they will not be} necessarily the same.
Note that we are not parameterizing $S(x)$ by the empirical value $\alpha_e$,
but with a new fitted value $\alpha_s$. This is in order to avoid biases, 
as a parameterization with $\alpha_e$ would lead to worse fits
(as the best one would be with $\alpha_s$) and therefore to larger values of
the resulting KS statistic and to artificially {larger} $p-$values.
So, although the null hypothesis of the test is that 
the exponent of the power law is $\alpha_e$,
and synthetic samples are obtained with this value, 
no further knowledge of this value is used in the test.
This is the procedure used by \citet{Clauset} and \citet{Malgrem},
but it is not clear if it is the one of \citet{Goldstein}.

In fact, one single synthetic sample is not enough to do a proper comparison
with the empirical sample, and we repeat the simulation many times
(avoiding of course to use the same seed of the random number generator for each sample).
The most important outcome is the set of values of the KS statistic, $d_s$,
which allows to estimate its distribution.
The $p-$value is simply calculated as 
\begin{equation}
p=\frac
{\mbox{number of simulations
with $d_s \ge d_e$}} {N_s},
\end{equation}
where $N_s$ is the number of simulations.
Figure 3 shows an example of the distribution of the KS statistic
for simulated data, which can be used as a table 
of critical values when the number of data and the exponent
are the same as in the example \cite{Goldstein}.

\begin{figure}[h]
\includegraphics*[height=10cm]
{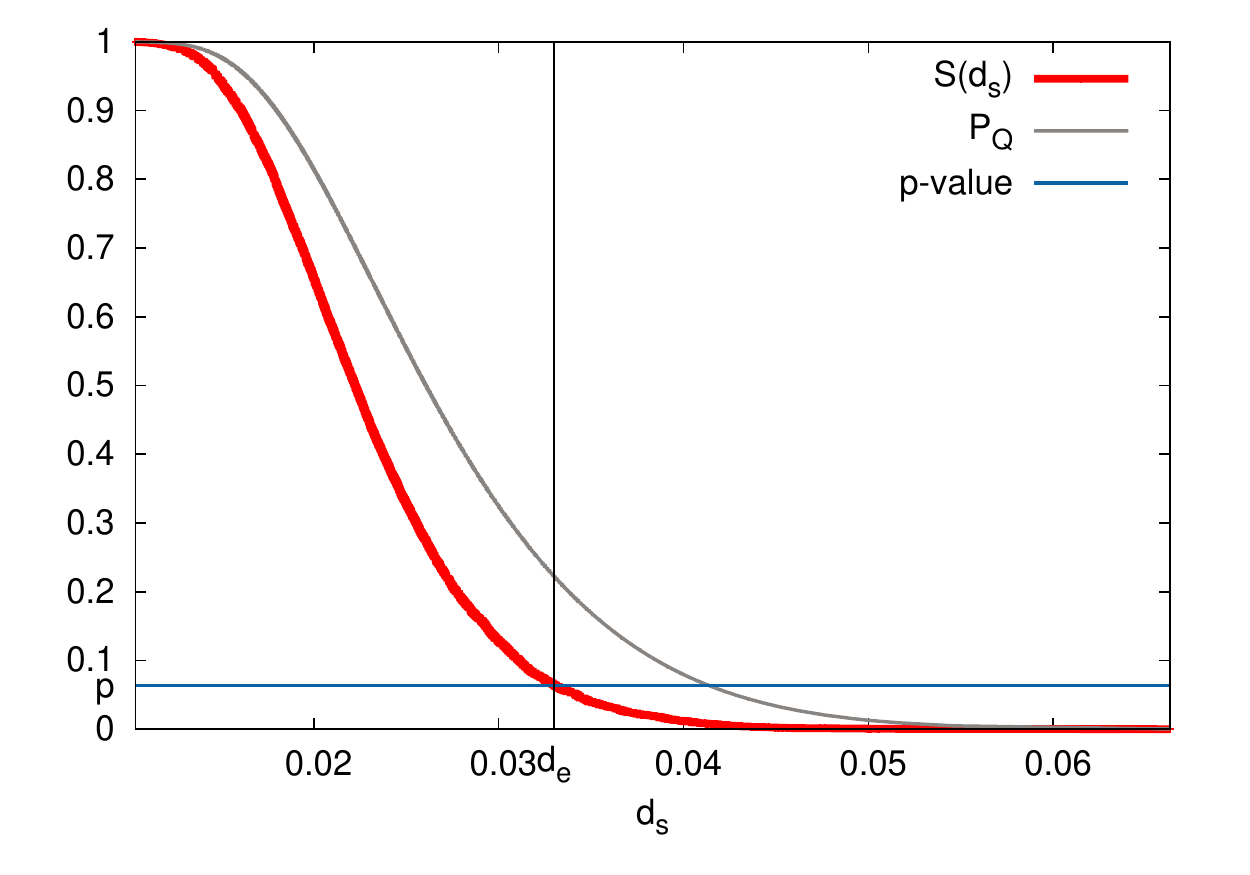}
\caption{Cumulative (complementary) distribution of the 
Kolmogorov-Smirnov statistic for simulated non-truncated power-law
distributions with $\alpha=\alpha_e=1.143$, $a=0.001$, and $N_{tot}=1000$.
The original ``empirical'' value $d_e=0.033$ is also shown.
The resulting $p-$value turns our to be $p=0.060 \pm 0.008$.
The ``false'' $p-$value, $p_Q$, arising from the KS formula, 
leads to higher values for the same $d_e$,
in concrete, $p_Q=0.22$.}  
\end{figure}

The standard deviation of the $p-$value can be calculated 
just using that the number of simulations with $d_s \ge d_e$
is binomially distributed, with standard deviation $\sqrt{N_s p (1-p)}$
and therefore the standard deviation of $p$ is the same divided by $N_s$,
\begin{equation}
\sigma_p=\sqrt{\frac{p(1-p)}{N_s}}.
\end{equation}
In fact, the $p-$value in this formula should be 
the ideal one (the one of the whole population)
but we need to replace it by the estimated 
value; further, when doing estimation from data, $N_s$
should be $N_s-1$, but we have disregarded
this bias correction. 
It will be also useful to consider the relative uncertainty
of $p$, which is the same as the relative uncertainty of 
the number of simulations with $d_s \ge d_e$
(as both are proportional).
Dividing the standard deviation of $p$ by its mean (which is $p$),
we obtain
\begin{equation}
CV_p=\sqrt{\frac{1-p}{p N_s}} \simeq \sqrt{\frac{1-p}{\mbox{number of simulations
with $d_s \ge d_e$}}}
\end{equation}
(we will recover this formula for the error of the estimation 
of the probability density).

In this way, small $p-$values are associated to large values of $d_e$,
and therefore to bad fits. However, note that if we put the threshold of rejection in,
let us say, $p \le 0.05$, 
even true power-law distributed data, with exponent $\alpha_e$, yield ``bad fits''
in one out of 20 samples (on average).
So we are rejecting true power laws in 5 \% of the cases (type I error).
On the other hand, lowering the threshold of rejection would reduce this problem, 
but would increase the probability of accepting false power laws (type II error).
In this type of tests a compromise between both 
types of errors is always necessary, 
and depends on the relative costs of rejecting a true hypothesis
or accepting a false one.

In addition, we can obtain from the Monte Carlo simulations
the uncertainty of the ML estimator, 
just computing $\bar \alpha_s$, the average value of $\alpha_s$,
and from here its standard deviation,
\begin{equation}
\sigma = \sqrt{\overline{(\alpha_s-\bar \alpha_s)^2}},
\end{equation}
{where the bars indicate average over the $N_s$ Monte Carlo simulations.}
This procedure yields good agreement with the analytical formula of 
\citet{Aban}, but can be much more useful in the discrete power-law
case. 

\subsection{Alternative method to the one by Clauset et al.}

At this point, for given values of the truncation points, $a$ and $b$,
we are able to obtain the corresponding ML estimation
of the power-law exponent as well as the goodness of the fit, 
by means of the $p-$value.
Now we face the same problem Clauset et al. (2009) tried to solve:
how to select 
{the fitting range?
In our case, how to find not only the value of $a$
but also of $b$?}
We adopt the simple method proposed by \citet{Peters_Deluca}:
sweeping many different values of $a$ and $b$ we should find, 
if the null hypothesis is true 
(i.e., if the sample is power-law distributed),
that many sets of intervals yield acceptable fits
(high enough $p-$values),
so we need to find the ``best'' of such intervals.
And which one is the best?
For a non-truncated power law the answer is easy, 
we select the largest interval, i.e., the one
with the smaller $a$,
{provided that the $p-$value is above
some fixed significance level $p_c$.}
All the other acceptable
intervals will be inside this one.

But if the power law is truncated the situation is 
not so clear, as there can be several non-overlapping
intervals. In fact, many true truncated power laws
can be contained in the data, 
at least there are well know examples of stochastic processes
with double power-law
distributions \cite{Levy,Boguna_Corral,Corral_pre.2003,Corral_jstat}.
At this point any selection can be reasonable, but if one insists
in having an automatic, blind procedure, a possibility is
to select either the interval which contains the larger
number of data, $N$ \cite{Peters_Deluca}, 
or the one which has the larger log-range, $b/a$.
For double power-law distributions, in which the exponent for small $x$
is smaller than the one for large $x$, the former recipe
has a tendency to select the first (small $x$) power-law regime,
whereas the second procedure 
{changes this tendency in some cases.}

In summary, the final step of the method for truncated power-law distributions 
is contained in the formula
\begin{equation}
[a,b]=\mbox{the one that yields higher }
\left\{
\begin{array}{c}
N\\
\mbox{or}\\
b/a\\
\end{array}
\right\}
\mbox{ provided that } p > p_c,
\end{equation}
which contains in fact two procedures, 
one maximizing $N$ and the other maximizing $b/a$.
We will test both in this paper.
For non-truncated power-law distributions the two procedures
are equivalent.


One might be tempted to choose $p_c=0.05$,
however, it is safer to consider a larger value, 
as for instance $p_c=0.20$.
Note that the $p-$value we are using is the one for 
fixed $a$ and $b$, and then the $p-$value
of the whole procedure should be different,
but at this point it is not necessary to obtain
such a $p-$value, 
as we should have already come out with a reasonable fit.
Figure 4 shows the results of the method
for true power-law data.

\begin{figure}[h]
\includegraphics*[height=13cm]
{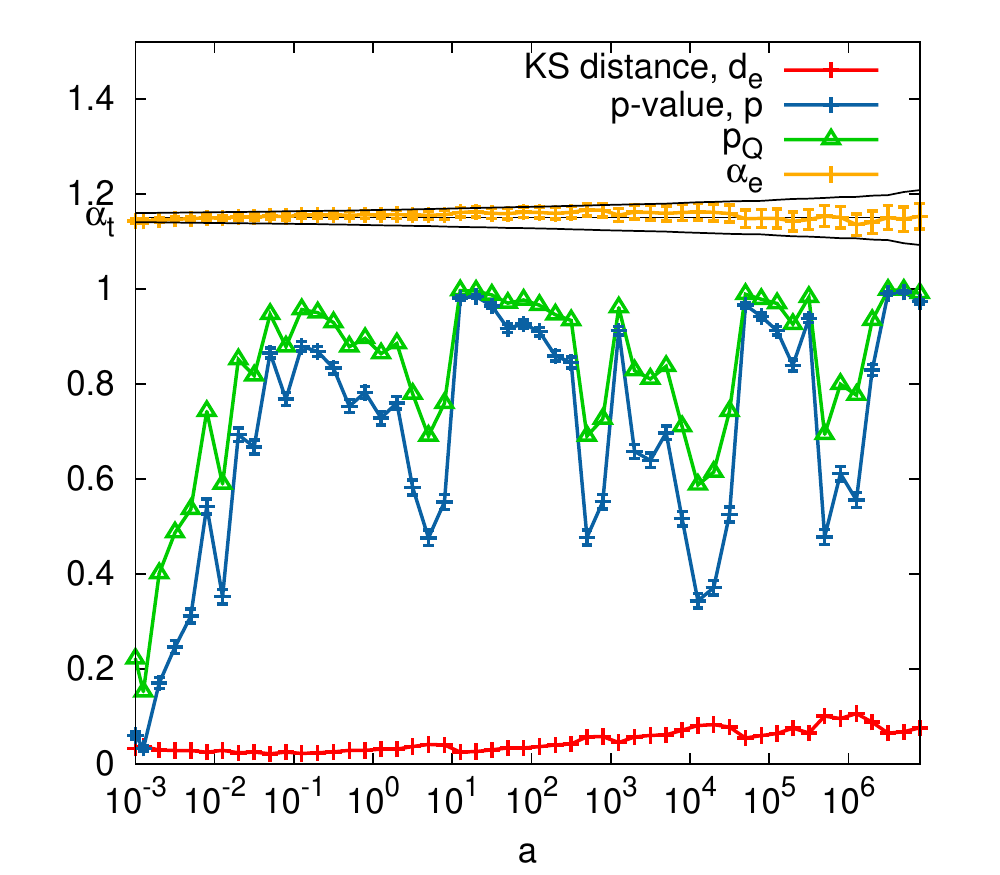}
\caption{Evolution as a function of $a$ 
of the KS statistic,
the false $p-$value $p_Q$,
the true $p-$ value (for fixed $a$),
and the estimated exponent.
The true exponent, 
here called $\alpha_t$ and
equal to 1.15, 
is displayed as a thin black line,
together with a $2\sigma$ interval.
Other horizontal lines mark the $p_c=0.20$ and $p_c=0.50$
limits of acceptance that select the resulting value of $a$.
}  
\end{figure}

\subsection{Truncated or non-truncated power-law distribution?}

{For broadly distributed data,}
the simplest choice is to try to fit 
first a non-truncated power-law distribution.
If an acceptable fit is found, it is expected that
a truncated power law, with $b \ge x_{max}$
({where $x_{max}$ is}
the largest value of $x$)
would yield also a good fit.
In fact, if $b$ is not considered as a fixed value but as a parameter
to fit, its maximum likelihood estimator when the number of data is fixed, 
i.e., when $b$ is in the range $b\ge x_{max}$, is $b_e=x_{max}$.
This is easy to see \cite{Aban}, just looking at the equations for $\ell(\alpha)$,
(11) and (12),
which show that $\ell(\alpha)$ increases as $b$ approaches $x_{max}$.
(In the same way, the ML estimator of $a$, for fixed number of data,
would be $a_e=x_{min}$, but we are not interested in such a case now.)
On the other hand, it is reasonable that a truncated power law yields a 
better fit than a non-truncated one, as the former has 
two parameters and the latter only one (assuming that $a$ is fixed, in any case).

In order to do a proper comparison, in such situations the so-called
Akaike information criterion (AIC) 
can be used. 
This is defined simply as the difference
between twice the number of parameters and
twice the maximum of the log-likelihood, i.e.,
\begin{equation}
AIC=2\times (\mbox{number of parameters})-2 \ell(\alpha_e).
\end{equation}
In general, having more parameters leads to better fits, 
and to higher likelihood, so, the first term
compensates this fact.
Therefore, given two models, the one with 
smaller AIC is preferred.
{Note that, in order that the comparison based on the AIC
makes sense, the fits that are compared have to be performed over
the same set of data. So, in our case this can only be done 
for non-truncated power laws and for truncated power laws with $b \ge x_{max}$.
Nevertheless, due to the limitations of this paper we have not performed the
comparison.}


\section{Estimation of probability densities and cumulative distribution functions}

The method of maximum likelihood does not rely on 
the estimation of the probability distributions,
in contrast to other methods.
Nevertheless, in order to present the results, 
it is useful to display some representation of the distribution, 
together with its fit.
This procedure has no statistical value
(it cannot provide a substitution of a goodness-of-fit test)
but is very helpful as a visual guide, 
specially in order to detect bugs in the algorithms.

\subsection{Estimation of the probability density}

{In the} definition of the probability density,
\begin{equation}
f(x) = \lim_{\Delta x \rightarrow 0}
\frac{\mbox{Prob}[x \le \mbox{random variable} < x+\Delta x]}{\Delta x},
\end{equation}
{a fundamental issue is that the width of the interval $\Delta x$ 
has to tend to zero.}
In practice $\Delta x$ cannot tend to zero 
(there would be no statistics in such case),
and one has to take a non-zero value of the width.
The most usual procedure is to draw a histogram using linear binning
(bins of constant width);
however, there is no reason why the width of the distribution
should be fixed (some authors even take $\Delta x=1$ as the {only possible choice}).
In fact, $\Delta x$ should be chosen in order to balance 
the necessity of having enough statistics (large $\Delta x$)
with that of having a good sampling of the function (small $\Delta x$).
For power-law distributions and other fat-tailed distributions,
which take values across many different scales, 
the right choice depends of the scale of $x$.
In this cases
it is very convenient
to use the so-called logarithmic binning \cite{Hergarten_book,Pruessner_book}.
This uses bins that appear as constant in logarithmic scale,
but that in fact grow exponentially (for which the method is
sometimes called exponential binning instead).
Curiously, this useful method is not considered by classic 
texts on density estimation
\cite{Silverman}.

Let us consider the semi-open intervals 
$[a_0,b_0), [a_1,b_1),\dots , [a_k,b_k), \dots$, also called bins, 
with $a_{k+1}=b_k$ and $b_k=B a_k$ 
(this constant $B$ has nothing to do with the one in the Gutenberg-Richter law, 
Sec. 1). For instance, if $B=\sqrt[5]{10}$ this yields 5 intervals
for each order of magnitude.
Notice
that the width of every bin grows linearly with $a_k$, 
but exponentially with $k$, as $b_k -a_k = (B-1)a_k= a_0(B-1)B^k$. 
The value of $B$ should be chosen in order to avoid low populated bins,
otherwise, a spurious exponent equal to one appears \cite{Pruessner_book}.

We simply will count the
number of occurrences of the variable in each bin. 
For each value of the random variable $x_i$, the corresponding bin is found as
\begin{equation}
k=\frac {\ln (x_i/a_0)}{\ln B}.
\end{equation}
Of course, $a_0$ has to be smaller than any possible value of $x$.
For a continuous variable the concrete value of $a_0$ should be irrelevant
(if it is small enough), but in practice one has to avoid that the values of
$a_k$ coincide with round values of the variable \cite{Corral_nuclear}.

So, with this logarithmic binning, the probability density
can be estimated (following its definition) as the relative frequency of
occurrences in a given bin divided by its width, i.e., 
\begin{equation}
f_e(x_k^*) = \frac{\mbox{number of occurrences in bin }k}{(b_k-a_k)\times 
\mbox{number of occurrences}},
\end{equation}
where the estimation of the density is associated to a value of $x$ 
represented by $x_k^*$.
The most practical solution is to take it in the middle of the
interval in logscale, so $x_k^*=\sqrt{a_k b_k}$.
However, for sparse data covering many orders of magnitude
it is necessary to be more careful.
In fact, what we are looking for is the
point $x_k^*$ whose value of the density
coincides with the probability of being between 
$a_k$ and $b_k$ divided by the with of the interval.
This is the solution of
\begin{equation}
f(x_k^*)=\frac 1 {b_k-a_k} \int_{a_k}^{b_k} f(x) dx = \frac {S(a_k)-S(b_k)}{b_k-a_k},
\end{equation}
where $f$ and $S$ are the theoretical distributions.
When the distribution and its parameters are known, 
the equation can be solved either analytically or numerically. 
It
is easy to see that for a power-law distribution (truncated or not) 
the solution can be written
\begin{equation}
x_k^*=\sqrt{a_k b_k} \left[(\alpha-1)\frac{B^{\alpha/2-1}(B-1)}{B^{\alpha-1}-1}\right]^{1/\alpha},
\end{equation}
where we have used that $B = b_k/a_k$ 
(if we were not using logarithmic binning we would have to
write a bin-dependent $B_k$). 
Note that for constant (bin-independent) $B$,
i.e., for logarithmic binning,
the solution is proportional but not equal to the geometric
mean of the extremes of the bin. Nevertheless, the omission of the proportionality factor does
not alter the power-law behavior, just shifts (in logarithmic scale) the curve. 
But for a different binning procedure
this is no longer true.
Moreover, for usual
values of $B$ the factor is very close to one \cite{Hergarten_book}, 
although large values of $B$ \cite{Corral_nuclear} yield noticeable
deviations if the factor in brackets is not included,
see also our treatment of the radionuclide half-lives in Sec. III,
with $B=1$.
Once the value of $B$ is fixed (usually in this paper to $\sqrt[5]{10}$), 
in order to avoid empty bins
we merge consecutive bins until the resulting
merged bins are not empty.
This leads to a change in the effective value of $B$
for merged bins,
but the method is still perfectly valid.

The uncertainty of $f_e(x)$ can be obtained from its standard deviation
(the standard deviation of the estimation of the density, $f_e$, not of 
the original random variable $x$).
Indeed, assuming independence in the sample
(which is already implicit in order to apply maximum likelihood estimation),
the number of occurrences of the variable in bin $k$ is a binomial random variable
(in the same way as for the $p-$value).
As the number of occurrences is proportional to $f_e(x)$,
the ratio between the standard deviation
and the mean for the number of occurrences will be the same
as for $f_e(x)$, which is,
\begin{equation}
\frac{\sigma_f (x)}{f_e(x)} = \sqrt{\frac {q}{\mbox{mean number of occurrences in $k$}}}
\simeq \frac{1}{\sqrt{\mbox{occurrences in $k$}}},
\end{equation}
where we replace the mean number of occurrences in bin $k$ (not available
from a finite sample) by the actual value, 
and $q$, the probability that the occurrences are not in bin $k$,
by one.
This estimation of $\sigma_f(x)$ fails when the number of counts
in the bin is too low, in particular if it is one.

One final consideration is that the fitted distributions 
are normalized between $a$ and $b$, with $N$ number of data,
whereas the empirical distributions include all data,
with $N_{tot}$ of them, $N_{tot}\ge N$.
Therefore, in order to compare the fits with the empirical distributions, 
we will plot $N f(x)/N_{tot}$ together with $f_e(x_k^*)$.

\subsection{Estimation of the cumulative distribution}

The estimation of the (complementary) cumulative distribution
is much simpler, as bins are not involved.
One just needs to sort the data, in ascending order, 
$x_{(1)} \le x_{(2)} \le \cdots \le x_{(N_{tot}-1)} \le x_{(N_{tot})}$;
then, the estimated cumulative distribution is
\begin{equation}
S_e(x_{(i)})=\frac{n_e(x_{(i)})}{N_{tot}}=\frac {N_{tot}-i+1} {N_{tot}},
\end{equation}
for the data points, constant below these data points, 
and $S_e(x)=0$ for $x>x_{(N_{tot})}$; 
$n_e(x_{(i)})$ is the number of data with $x\ge x_{(i)}$
in the empirical sample.
We use the case of empirical data as an example,
but it is of course the same for simulated data.
For the comparison of the empirical distribution with the theoretical 
fit we need to correct the different number of data in both cases.
So, we plot both
$[N S(x)+n_e(b)]/N_{tot}$ and $S_e(x)$, 
in order to check the accuracy of the fit.

\section{Data Analyzed and Results}

We have explained how, in order to try to certify 
that a set of data is compatible with a simple power-law distribution,
a large number of mathematical formulas and immense calculations
are required.
Now we check the performance of our method with diverse geophysical data,
which were previously analyzed with different, less rigorous or
worse-functioning methods.
For the peculiarities and challenges of the data set, 
we also include the half-lives of unstable nuclides.
The parameters of the method are fixed to $N_s=1000$ Monte Carlo simulations
and the values of $a$ and $b$ are found sweeping 
a fixed number of points per order of magnitude,
equally spaced in logarithmic scale.
This number is 
10 for non-truncated power laws (in which $b$ is fixed to infinity)
and 5 for truncated power laws.
Three values of $p_c$ are considered: 0.1, 0.2, and 0.5,
in order to compare the dependence of the results on this parameter.
The results are reported using the Kolmogorov-Smirnov test for goodness-of-fit.
If, instead, the Kuiper's test is used, 
the outcome is not significantly different in most of the cases.
In a few cases the fitting range, and therefore the exponent, changes,
but without a clear trend, i.e., the fitting range can become smaller
or increase. These cases deserve a more in-depth investigation.

\subsection{Half-lives of the radioactive elements}

\citet{Corral_nuclear} 
studied the statistics of the half-lives of radionuclides
(comprising both nuclei in the fundamental and in excited states).
Any radionuclide has a constant probability of disintegration per unit time, 
the decay constant,
let us call it $\lambda$ \cite{Krane}. 
If $M$ is the total amount of radioactive material at
time $t$, this means that
\begin{equation}
-\frac 1 {M} \frac {dM}{dt} = \lambda.
\end{equation}
This leads to an exponential decay, 
for which a half-life $t_{1/2}$ or a lifetime $\tau$ can be defined, as
\begin{equation}
t_{1/2}= \tau \ln 2= \frac{\ln 2}{\lambda}.
\end{equation}
It is well known that the half-lives take disparate values, for example, that of $^{238}$U
is 4.47 (American) billions of years, whereas for other nuclides 
it is a very tiny fraction of a second.

It has been recently claimed that these half-lives are power-law distributed
\cite{Corral_nuclear}.
In fact, three power-law regions were identified in the probability density of $t_{1/2}$,
roughly,
\begin{equation}
f(t_{1/2})\propto\left\{
\begin{array}{ll}
1 / {t_{1/2}^{0.65}} & \mbox{ for } 10^{-6} s \le t_{1/2} \le 0.1 s\\
1 / {t_{1/2}^{1.19}} & \mbox{ for } 100 s \le t_{1/2} \le 10^{10} s\\
1 / {t_{1/2}^{1.09}} & \mbox{ for } t_{1/2} \ge  10^{8} s.\\
\end{array}
\right.
\end{equation}
Notice that there is some overlap between two of the intervals,
as reported in the original reference, due to problems 
in delimiting the transition region.
The study used variations of the Clauset et al.'s (2009) method
of minimization of the KS statistic,
introducing and upper cutoff and additional conditions to escape from the
global minimum of the KS statistic,
which yielded the rejection ($p=0.000$) of the power-law hypothesis.
These additional conditions were of the type of taking either $a$ or 
$b/a $ greater than a fixed amount.

For comparison, we will apply the method explained in the previous section to this problem.
Obviously, our random variable will be $x=t_{1/2}$.
The data is exactly the same as in the original reference, 
coming from the Lund/LBNL Nuclear
Data Search web page \cite{Firestone}.
Elements whose half-life is only either bounded from below
or from above are discarded for the study, 
which leads to 3002 radionuclides with well-defined
half-lives; 2279 of them are in their ground state
and the remaining 723 in an exited state.
The minimum and maximum half-lives in the data set 
are 
$3 \times 10^{-22}$ s and
$7\times 10^{31}$ s, respectively,
yielding more than 53 orders of magnitude of logarithmic range.
Further details are in \citet{Corral_nuclear}.

The results of our fitting and testing method 
are shown in Table I and in Fig. 5.
The fitting of a non-truncated power law 
yields results in agreement with \citet{Corral_nuclear},
with $\alpha=1.09 \pm 0.01$ and $a=3\times 10^7$ s,
for the three values of $p_c$ analyzed (0.1, 0.2, and 0.5).
When fitting a truncated power law,
the maximization of the log-range, $b/a$,
yields essentially the same results as
for a non-truncated power law,
with slightly smaller exponents $\alpha$
due to the finiteness of $b$ (results not shown).
In contrast, the maximization of the number of data $N$
yields an exponent $\alpha\simeq 0.95$ between $a\simeq 0.1$ s
and $b \simeq 400$ s (with some variations depending on $p_c$).
This result is in disagreement with \citet{Corral_nuclear},
which yielded a smaller exponent for smaller values of $a$ and $b$.
In fact, as the intervals do not overlap both results are compatible, 
but it is also likely that a different function would lead to 
a better fit; for instance, a lognormal 
between 0.01 s a $10^5$ s was proposed by \citet{Corral_nuclear},
although the fitting procedure there was not totally reliable.
Finally, the intermediate power-law range reported in the original
paper (the one with $\alpha=1.19$) is not found by any of our algorithms
working on the entire dataset.
It is necessary to cut the data set, removing data below, 
for instance, 100 s (which is equivalent to impose $a > 100$ s),
in order that the algorithm converges to that solution.
So, caution must be taken when applying the algorithm blindly, 
as important power-law regimes may be hidden by others having either
larger $N$ or larger log-range.




\begin{table*}[t]
 \caption{ 
Results of the fits for the $N_{tot}=3002$ nuclide half-lives data,
for different values of $p_c$.
We show the cases of a pure or non-truncated power law (with $b=\infty$, fixed)
and truncated power law (with $b$ finite, estimated from data), maximizing $N$.
{The latter is splitted into two subcases: exploring the whole range of $a$
(rows 4,5, and 6) and restricting $a$ to $a > 100$ s (rows 7, 8, and 9).}
\label{results}}
\centering
\begin{tabular}{r c c c r c c}
 $N$  &  $a$ (s) & $b$ (s) & ${b}/{a}$ & $\alpha \pm \sigma$ & $p_c$\\ 
\hline 
     143 &       0.316 $\times 10^8$
&       $\infty$  &       $\infty$  &  1.089$\pm$ 0.007   
& 0.10 \\ 
    143 &       0.316 $\times 10^8$
 &       $\infty$ &       $\infty$  &  1.089$\pm$ 0.007 
 & 0.20  \\ 
    143 &       0.316 $\times 10^8$
&       $\infty$ &       $\infty$  &  1.089$\pm$ 0.008  
& 0.50     \\ 
\hline
    1596 &               0.0794 &       501 &       6310 &  0.952$\pm$ 0.010    & 0.10\\ 
    1539 &               0.1259 &       501 &       3981 &  0.959$\pm$ 0.011   & 0.20 \\ 
    1458 &               0.1259 &       316 &       2512 &  0.950$\pm$ 0.011   & 0.50 \\ 
\hline
 1311 &   
 125.9 &       0.501 $\times 10^{23}$ &       0.398 $\times 10^{21}$ &  1.172$\pm$ 0.005  & 0.10   \\ 
 1309 &   
 125.9 &       0.316 $\times 10^{22}$ &       0.251 $\times 10^{20}$ &  1.175$\pm$ 0.005   & 0.20  \\ 
 1303 &   
 125.9 &       0.794 $\times 10^{18}$ &       0.631 $\times 10^{16}$ &  1.177$\pm$ 0.005    & 0.50 \\ 
\hline
\end{tabular}
\end{table*}


\begin{figure}[h]
\includegraphics*[height=10cm]{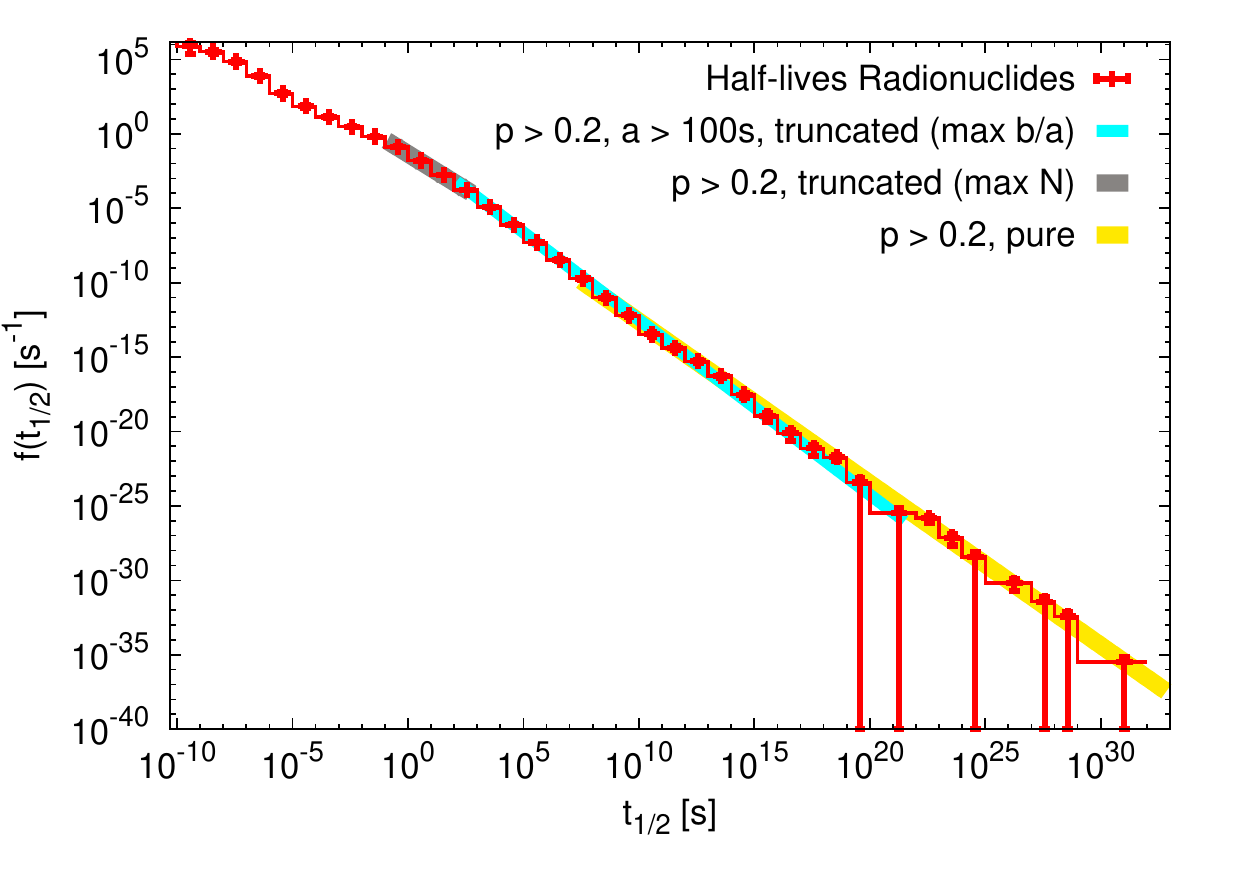}
\caption{
Estimated probability density of the half-lives of the radionuclides,
together with the power-law fits explained in the text.
The number of log-bins per order of magnitude is one, 
which poses a challenge in the correct estimation of the exponent, 
as explained in Sec. II.
Data below $10^{-10}$ s are not shown.
}  
\end{figure}
%

\subsection{Seismic moment of earthquakes}

The statistics of the sizes of earthquakes 
\cite{Gutenberg_Richter}
has been investigated
not only since the introduction of the first magnitude scale, 
by C. Richter, but even before, in the 1930's, by Wadati
\cite{Utsu_GR}.
From a modern perspective, the most reliable measure of earthquake
size is given by the (scalar) seismic moment $M$,
which is the product of mean final slip, rupture area, and the rigidity of the fault material
\cite{Ben_Zion_review}.
It is usually assumed that the energy radiated by an earthquake is proportional to 
the seismic moment \cite{Kanamori_rpp},
so, a power-law distribution of the seismic moment implies a power-law distribution
of energies, with the same exponent.

The most relevant results for the distribution of seismic moment are those 
of Kagan for worldwide seismicity \cite{Kagan_gji02}, 
who showed that its probability density has a power-law body, 
with a universal exponent in agreement with $\alpha=1.63 \simeq 5/3$,
but with an extra, non-universal exponential decay
(at least in terms of the complementary cumulative distribution).
However, Kagan's (2002) analysis, ending in 2000, 
refers to a period of global seismic ``quiescence'';
in particular, the large Sumatra-Andaman earthquake of 2004
and the subsequent global increase of seismic activity
are not included.
Much recently, \citet{Main_ng} have shown, using a Bayesian information criterion, 
that the inclusion of the new events leads to the preference of the non-truncated
power-law distribution in front of models with a faster large-$M$ decay.

We take the Centroid Moment Tensor (CMT) worldwide catalog analyzed by Kagan (2002)
and by Main et al. (2008), including now data from January 1977 to December 2010,
and apply our statistical method to it.
Although the statistical analysis of Kagan is rather complete, 
his procedure is different to ours.
Note also that the data set does not comprise the recent (2011) Tohoku earthquake in Japan,
nevertheless, the qualitative change in the data with respect to Kagan's period of analysis 
is very remarkable.
Following this author, we separate the events by their depth:
shallow for depth $\le 70$ km,
intermediate for 70 km $<$ depth $\le 300$ km,
and
deep for depth $>$ 300 km.
The number of earthquakes in each category
is 26824, 5281, and 1659, respectively.

Second, we also consider the Southern California's
Waveform 
Relocated Earthquake Catalog, 
from 
January 1st, 1981 to June 30th, 2011,
covering 
 a rectangular box of coordinates ($122^\circ$W,$30^\circ$N), ($113^\circ$W,$37.5^\circ$N)
\cite{Shearer2,Shearer3}. 
This catalog contains 111981 events with $m\ge 2$.
As, in contrast with the CMT catalog, this one 
does not report the seismic moment $M$,
the magnitudes $m$ there are converted into seismic moments,
using the formula
\begin{equation}
\log_{10} M = \frac 3 2 (m +6.07),
\end{equation}
where $M$ comes in units of Nm (Newtons times meters);
however, this formula is a very rough estimation
of seismic moment, as it is only accurate (and exact) when $m$
is the so-called moment magnitude \cite{Kanamori_rpp}, 
whereas the magnitudes recorded in the catalog
are of a different type.
In any case, our procedure here is equivalent 
to fit an exponential distribution to the 
magnitudes reported in the catalog.

Tables II and III and Fig. 6
summarize the results of analyzing these data
with our method, taking $x=M$ as the random variable.
Starting with the non-truncated power-law distribution,
we always obtain an acceptable (in the sense of non-rejectable) 
power-law fit, valid for
several orders of magnitude.
In all cases the exponent $\alpha$ is between 1.61 and 1.71,
but for Southern California it is always very close to 1.66.
For the worldwide CMT data the largest value of $a$ is $3 \times 10^{18}$ Nm,
corresponding to a magnitude $m=6.25$ (for shallow depth),
and the smallest is $a=8 \times 10^{16}$ Nm,
corresponding to $m=5.2$ (intermediate depth).
If the events are not separated in terms of their depth
(not shown), 
the results are dominated by the shallow case,
except for $p_c=0.5$, which leads to very large values of
$a$ and $\alpha$ ($a=5\times 10^{20}$ Nm and $\alpha\simeq 2$).
The reason is probably the mixture of the different
populations, in terms of depth, which is not recommended by Kagan (2002).
This is an indication that the inclusion of an upper limit $b$
to the power law may be appropriate, with each depth corresponding to 
different $b$'s.
For Southern California, the largest $a$ found (for $p_c=0.5$)
is $1.6 \times 10^{15}$ Nm,
giving $m=4$.
This value is somewhat higher, in comparison with the completeness magnitude
of the catalog; perhaps the reason that the power-law fit is rejected
for smaller magnitudes is due to the fact that these magnitudes are
not true moment magnitudes, but come from a mixture of different
magnitude definitions.
If the value of $a$ is increased, the number of data $N$ is decreased
and the power-law hypothesis is more difficult to reject,
due simply to poorer statistics.
When a truncated power law is fitted, using the method 
of maximizing the number of data leads to similar values 
of the exponents, although the range of the fit is
in some cases moved to smaller values (smaller $a$, 
and $b$ smaller than the maximum $M$ on the dataset).
The method of maximizing $b/a$ leads to results
that are very close to the non-truncated power law.


\begin{table*}[t]
 \caption{ 
Results of the non-truncated power-law fit ($b=\infty$) applied to the
seismic moment of earthquakes in
CMT worldwide catalog (separating by depth) and to the
Southern California catalog, for different $p_c$.
}
\centering
\begin{tabular}{l r c c c c }
{Catalog} & $N$  &  $a$  & $\alpha \pm \sigma$ & $p_c$\\ 

\hline 
CMT deep &    1216 &     0.1259 $\times 10^{18}$ 
&  1.622$\pm$ 0.019    & 0.10 \\ 
intermediate &    3701 &     0.7943 $\times 10^{17}$ 
&  1.654$\pm$ 0.011    & 0.10 \\ 
shallow &    5799 &    0.5012 $\times 10^{18}$ 
&  1.681$\pm$ 0.009    & 0.10 \\ 
\hline
CMT deep &     898 &     0.1995 $\times 10^{18}$ 
&  1.608$\pm$ 0.020    & 0.20 \\ 
intermediate &    3701 &     0.7943 $\times 10^{17}$ 
&  1.654$\pm$ 0.011    & 0.20 \\ 
shallow &    5799 &    0.5012 $\times 10^{18}$ 
&  1.681$\pm$ 0.009    & 0.20 \\ 
\hline
CMT deep &     898 &     0.1995 $\times 10^{18}$ 
 &  1.608$\pm$ 0.021    & 0.50 \\ 
intermediate &    3701 &    0.7943 $\times 10^{17}$ 
&  1.654$\pm$ 0.011    & 0.50 \\ 
shallow &    1689 &   0.3162 $\times 10^{19}$ 
&  1.706$\pm$ 0.018    & 0.50 \\ 
\hline
\hline
S. California &     1327 &0.1000 $\times 10^{16}$ 
&  1.660$\pm$ 0.018 & 0.10    \\ 
S. California &     1327 &0.1000 $\times 10^{16}$ 
&  1.660$\pm$ 0.018 & 0.20    \\ 
S. California &     972  &0.1585 $\times 10^{16}$ 
&  1.654$\pm$ 0.021 & 0.50    \\ 
\hline
\end{tabular}
\end{table*}

\begin{table*}[t]
 \caption{ 
Same as the previous table, but fitting a truncated power-law distribution, 
by maximizing the number of data.
}
\centering
\begin{tabular}{l r c c c c c c }
{Catalog} & $N$  & $a$ & $b$ & ${b}/{a}$ & $\alpha \pm \sigma$ & $p_c$\\ 
\hline 
CMT deep &    1216 &     0.1259 $\times 10^{18}$ &       0.3162 $\times 10^{23}$ &       0.2512 $\times 10^6$ &  1.621$\pm$ 0.019    & 0.10\\ 
intermediate &    3701 &    0.7943 $\times 10^{17}$ &       0.7943 $\times 10^{22}$ &       0.1000 $\times 10^6$ &  1.655$\pm$ 0.011    & 0.10\\ 
shallow &   13740 &    0.1259 $\times 10^{18}$ &       0.5012 $\times 10^{20}$ &       0.3981 $\times 10^3$ &  1.642$\pm$ 0.007    & 0.10\\ 
\hline
CMT deep &    1076 &     0.1259 $\times 10^{18}$ &       0.5012 $\times 10^{19}$ &       0.3981 $\times 10^2$ &  1.674$\pm$ 0.033    & 0.20\\ 
intermediate &    3701 &     0.7943 $\times 10^{17}$ &       0.7943 $\times 10^{22}$ &       0.1000 $\times 10^6$ &  1.655$\pm$ 0.011    & 0.20\\ 
shallow &   13518 &    0.1259 $\times 10^{18}$ &       0.1995 $\times 10^{20}$ &       0.1585 $\times 10^3$ &  1.636$\pm$ 0.008    & 0.20\\ 
\hline
CMT deep &     898 &     0.1995 $\times 10^{18}$ &       0.3162 $\times 10^{23}$ &       0.1585 $\times 10^6$ &  1.604$\pm$ 0.021    & 0.50\\ 
intermediate &    3701 &     0.7943 $\times 10^{17}$ &       0.7943 $\times 10^{22}$ &       0.1000 $\times 10^6$ &  1.655$\pm$ 0.011    & 0.50\\ 
shallow &   11727 &    0.1259 $\times 10^{18}$ &       0.1995 $\times 10^{19}$ &       0.1585 $\times 10^2$ &  1.608$\pm$ 0.012    & 0.50\\ 
\hline
\hline
S. California &     1146 &0.1259 $\times 10^{16}$ &       0.1259 $\times 10^{22}$ &   0.1000 $\times 10^7$
&  1.663$\pm$ 0.020 & 0.10    \\ 
S. California &     1146 &0.1259 $\times 10^{16}$ &       0.1259 $\times 10^{22}$ &   0.1000 $\times 10^7$
&  1.663$\pm$ 0.020 & 0.20    \\ 
S. California &     344 &0.7943 $\times 10^{16}$ &       0.1259 $\times 10^{22}$ &   0.1585 $\times 10^6$
&  1.664$\pm$ 0.036 & 0.50    \\ 
\hline
\end{tabular}
\end{table*}


\begin{figure}[h]
\includegraphics*[height=12cm,angle=0]{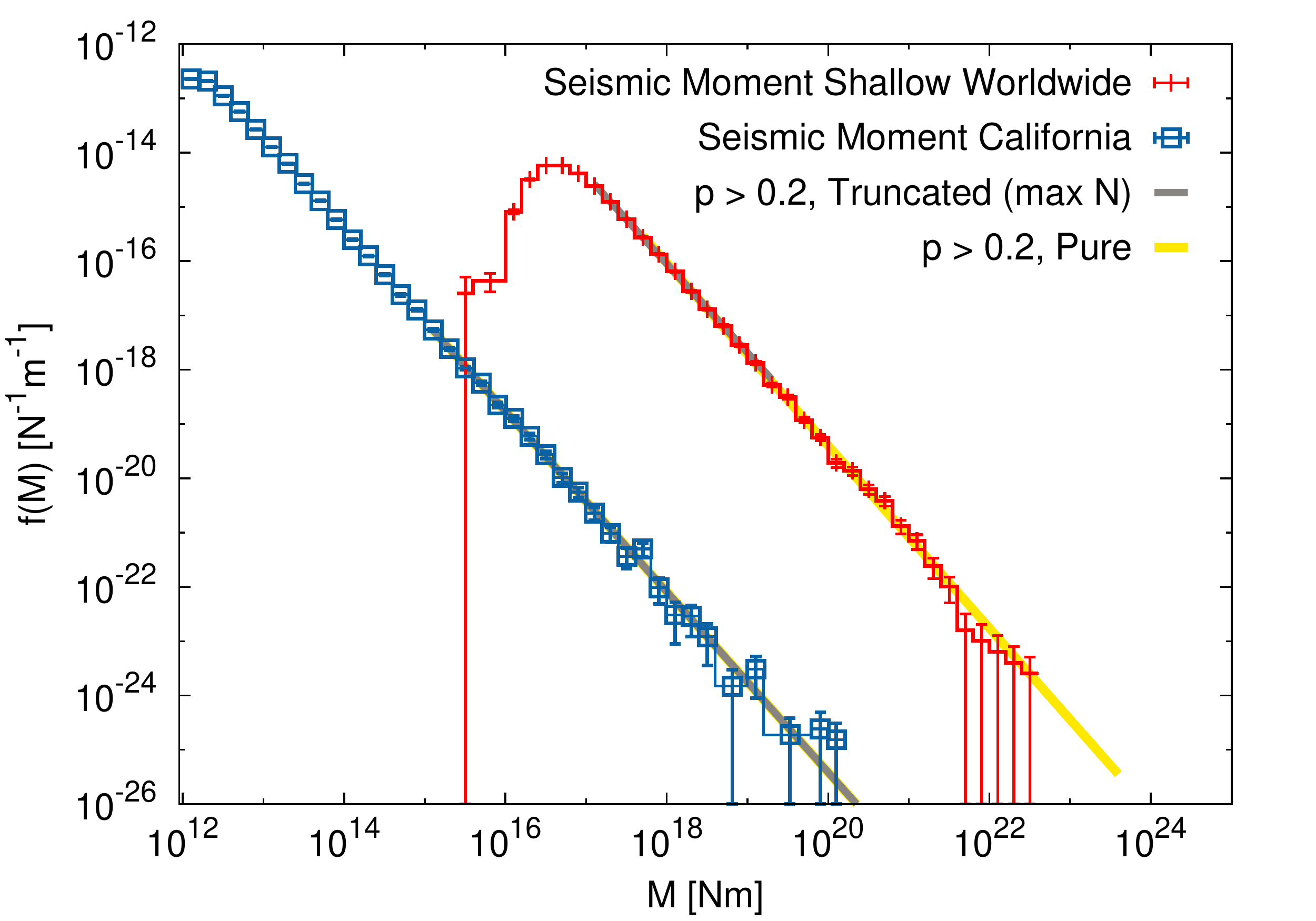}
\caption{Estimated probability densities and corresponding power-law fits
of the seismic moment $M$ of shallow earthquakes in the worldwide CMT catalog
and of the estimated $M$ in the Southern California catalog.}  
\end{figure}

\subsection{Energy of tropical cyclones}

Tropical cyclones are devastating atmospheric-oceanic
phenomena comprising tropical depressions, tropical storms,
and hurricanes or typhoons \cite{Emanuel_book}.
Although the counts of events every year have been monitored
for a long time, and other measurements to evaluate
annual activity have been introduced
(see \citet{Corral_agu} for an overview), little attention 
has been paid to the statistics of individual tropical cyclones.

In 2005,
Emanuel introduced the power-dissipation index (PDI) 
as a simple way to obtain a rough estimation of the
total energy dissipated by all tropical cyclones
in a given season and some ocean basin \cite{Emanuel_nature05}.
But the PDI can also be used to characterize individual 
events as well, as it was done later by \citet{Corral_hurricanes}.
Indeed, the PDI is defined as the sum for all the discrete times $t$
(that comprise the lifetime of a tropical cyclone) of the cube 
of the maximum sustained wind speed multiplied by the 
time interval of sampling, $\Delta t$.
In a formula, 
\begin{equation}
PDI= \sum_{\forall t} v_t^3 \Delta t,
\end{equation}
where $v_t$ is the maximum sustained wind speed.
In the so-called best-track records, $\Delta t = 6$ hours;
this factor would only be necessary in order to compare
with other data with different $\Delta t$
(but caution should be present in this case for the
possible fractal nature of the speed signal).
Although the speeds are reported in knots, they are converted
to m/s (using that 1 knot$=$0.514 m/s), and then we report the PDI in m$^3$/s$^2$.

\citet{Corral_hurricanes} studied the statistics 
of the PDI (defined for individual events, in contrast to 
Emanuel's (2005b) work) in 4 different ocean basins for several 
time periods. 
The results showed a rapid, perhaps exponential, decay at the tail,
but a body of the distribution compatible with a power law, 
for 1 or 2 orders of magnitude,
with exponents close to one.
The connection with SOC phenomena
was discussed by \citet{Corral_Elsner}.
The method used was again a variation of the Clauset et al.'s (2009) one,
introducing an upper cutoff and additional restrictions to the 
variations of the parameters.
Here we revisit this problem, trying to use updated data (whenever it
has been possible), and applying the method which is the subject of this paper
to $x=PDI$.

The data has been downloaded from the National Hurricane Center (NHC) of NOAA,
for the North Atlantic and the Northeastern Pacific \cite{NOAA,NOAA_hurdat2}
and from the Joint Typhoon Warning Center (JTWC) of the US Navy \cite{ATCR_report,ATCR2}
for the Northwestern Pacific, the Southern Hemisphere
(comprising the Southern Indian and the Southwestern Pacific),
and the Northern Indian Ocean.
The abbreviation, time span, and number of events
for each basin are:
NAtl, 1966--2011, 532;
EPac, 1966--2011, 728;
WPac, 1986--2011, 690;
SHem, 1986--2007 (up to May), 523; 
NInd, 1986--2007, 110.
The latter case was not studied in 
any of the previous works.

The results for a truncated power law maximizing $N$, shown in Table IV and Fig. 7,
are in agreement with those of \citet{Corral_hurricanes}.
In general, exponents are close but above 1,
except for the Northwestern Pacific, where $\alpha \simeq 0.96$,
and for the North Indian,
where $\alpha$ is substantially higher than one.
We consider that this method performs rather well.
It would be interesting to test if universality can nevertheless hold
(the high value for the North Indian Ocean is based in 
much less data than for the rest of basins),
or if there is some systematic bias in the value of the exponents
(the protocols of the NHC and the JTWC are different, 
and the satellite coverage of each basin is also different).

\begin{figure}[h]
\includegraphics*[height=18cm]
{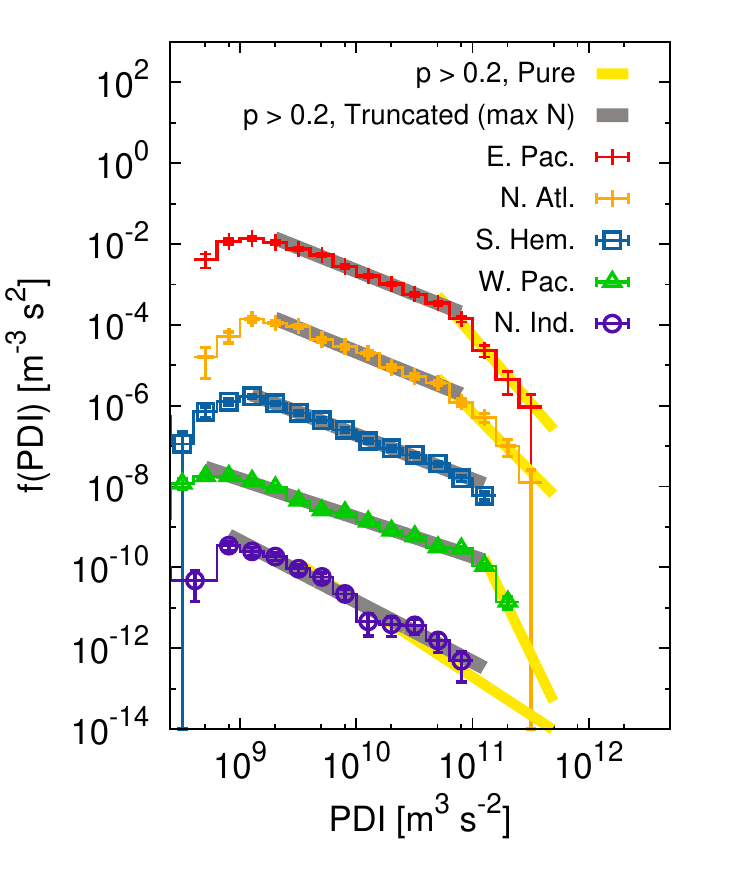}
\caption{Estimated probability densities of the PDI of tropical cyclones
in 5 ocean basins, together with their power-law fits.
The values of the densities are multiplied by $1$,
$10^{2}$, $10^{4}$, $10^{6}$, and $10^{8}$, 
for clarity sake.}  
\end{figure}

If a non-truncated power law is fit to the data,
the fits turn out to be rather short, with a high exponent (up to 6)
describing the tail of the distribution
(except for the Southern Hemisphere, where no such tail
is apparent).
We do not give any relevance to these results,
as other alternatives, as for instance a simple exponential
tail, have to be considered \cite{Corral_agu,Castillo_Serra}.
Coming back to a truncated power law, but maximizing the log-range,
the algorithm sometimes fits the power law in the body of the distribution
(with exponent close to 1) and for some other times
the algorithm goes to the fast-decaying tail.
So the method of maximizing $b/a$ is not useful for this data.

\begin{table*}[t]
 \caption{ 
Results of the truncated power-law fits, maximizing $N$, 
for the PDI in the 5 ocean basins with tropical-cyclone activity,
for different values of $p_c$.
\label{results2}}
\centering
\begin{tabular}{l r c c r c c }
basin & $N$   & $a$ (m$^3/$s$^2$) & $b$ (m$^3/$s$^2$)& ${b}/{a}$ & $\alpha \pm \sigma$ & $p_c$\\ 
\hline 
EPac&    637 &      0.1259 $\times 10^{10}$ &       0.7943 $\times 10^{11}$ &       63 &  1.094$\pm$ 0.033    & 0.10\\ 
NAtl&    417 &       0.1995 $\times 10^{10}$ &       0.7943 $\times 10^{11}$ &       40 &  1.168$\pm$ 0.047    & 0.10\\ 
SHem&     523 &       0.1259 $\times 10^{10}$ &       0.1259 $\times 10^{12}$ &       100 &  1.108$\pm$ 0.034    & 0.10\\ 
WPac&    637 &         0.5012 $\times 10^9$ &       0.1259 $\times 10^{12}$ &       251 &  0.957$\pm$ 0.025    & 0.10\\ 
NInd&     102 &        0.7943 $\times 10^{9\phantom{1}}$ &       0.1995 $\times 10^{12}$ &       251 &  1.520$\pm$ 0.077    & 0.10\\ 
\hline
EPac&      571 &       0.1995 $\times 10^{10}$ &       0.7943 $\times 10^{11}$ &       40 &  1.149$\pm$ 0.039    & 0.20\\ 
NAtl&     417 &       0.1995 $\times 10^{10}$ &       0.7943 $\times 10^{11}$ &       40 &  1.168$\pm$ 0.047    & 0.20\\ 
SHem&    523 &       0.1259 $\times 10^{10}$ &       0.1259 $\times 10^{12}$ &       100 &  1.108$\pm$ 0.033    & 0.20\\ 
WPac&    637 &        0.5012 $\times 10^{9\phantom{1}}$ &       0.1259 $\times 10^{12}$ &       251 &  0.957$\pm$ 0.025    & 0.20\\ 
NInd&    102 &        0.7943 $\times 10^{9\phantom{1}}$ &       0.1259 $\times 10^{12}$ &       158 &  1.490$\pm$ 0.077    & 0.20\\ 
\hline
EPac&      571 &       0.1995 $\times 10^{10}$ &       0.7943 $\times 10^{11}$ &       40 &  1.149$\pm$ 0.040    & 0.50\\ 
NAtl&     417 &      0.1995 $\times 10^{10}$ &       0.7943 $\times 10^{11}$ &       40 &  1.168$\pm$ 0.045    & 0.50\\ 
SHem&465 &       0.1995 $\times 10^{10}$ &       0.1259 $\times 10^{12}$ &       63 &  1.132$\pm$ 0.040    & 0.50\\ 
WPac&     637 &        0.5012 $\times 10^{9\phantom{1}}$ &       0.1259 $\times 10^{12}$ &       251 &  0.957$\pm$ 0.024    & 0.50\\ 
NInd&    86 &        0.7943 $\times 10^{9\phantom{1}}$ &       0.1259 $\times 10^{11}$ &       16 &  1.323$\pm$ 0.139    & 0.50\\ 

\hline
\end{tabular}
\end{table*}
\subsection{Area of forest fires}

The statistics of the size of forest fires was an intense topic of research
since the introduction of the concept of SOC, at the end of the
1980's, but only from the point of view of cellular-automaton models.
Real data analysis had to wait several years
\cite{Malamud_science,Malamud_fires_pnas},
leading to power-law distributions, 
more or less in agreement with the models.
Here we are particularly interested in a dataset
from Italy, for which a power-law distribution
of sizes was ruled out \cite{Corral_fires}.
Instead, a lognormal tail was proposed
for the fire size probability density.

The data considered by \citet{Corral_fires},
and reanalyzed in this study, comes from
the Archivio Incendi Boschivi (AIB) fire catalog compiled
by the (Italian) 
\citet{Italy_fire_catalog}.
The subcatalog to which we restrict covers all Italy and spans
the 5-year period 1998-2002, containing
36 748 fires. 
The size of each fire is measured by
the burned area $A$, in hectares, with 1 ha=$10^4$ m$^2$.
In this subsection we analyze the case of $x=A$.

The results in Table V and Fig. 8 show that a pure (non-truncated)
power law is only acceptable (in the sense of non-rejectable)
for the rightmost part of the tail of the distribution,
comprising less than one order of magnitude.
It is very indicative that only 51 data are in the 
possible power-law tail.
Therefore, we disregard this power-law behavior as 
spurious and expect that other distributions
can yield a much better fit
(not in order of the quality of the fit but regarding
the number of data it spans).
This seems in agreement with other analyses of 
forest-fire data \cite{Newman_05,Clauset}.
If a truncated power-law is considered, fitted by maximizing
the number of data, 
the results are not clearly better, as seen in the table.
Moreover, there is considerable variation with the 
value of $p_c$. So, we do not give any relevance to such power-law fits.
Finally, the method of maximizing $b/a$ yields the same
results as for the non-truncated power law
(except by the fact that the exponents are slightly smaller,
not shown).
In order to provide some evidence for the better adequacy of the
lognormal tail in front of the power-law tail for these data, 
it would be interesting to apply an adaptation of the test
explained by \citet{Castillo}.

\begin{figure}[h]
\includegraphics*[height=10cm]
{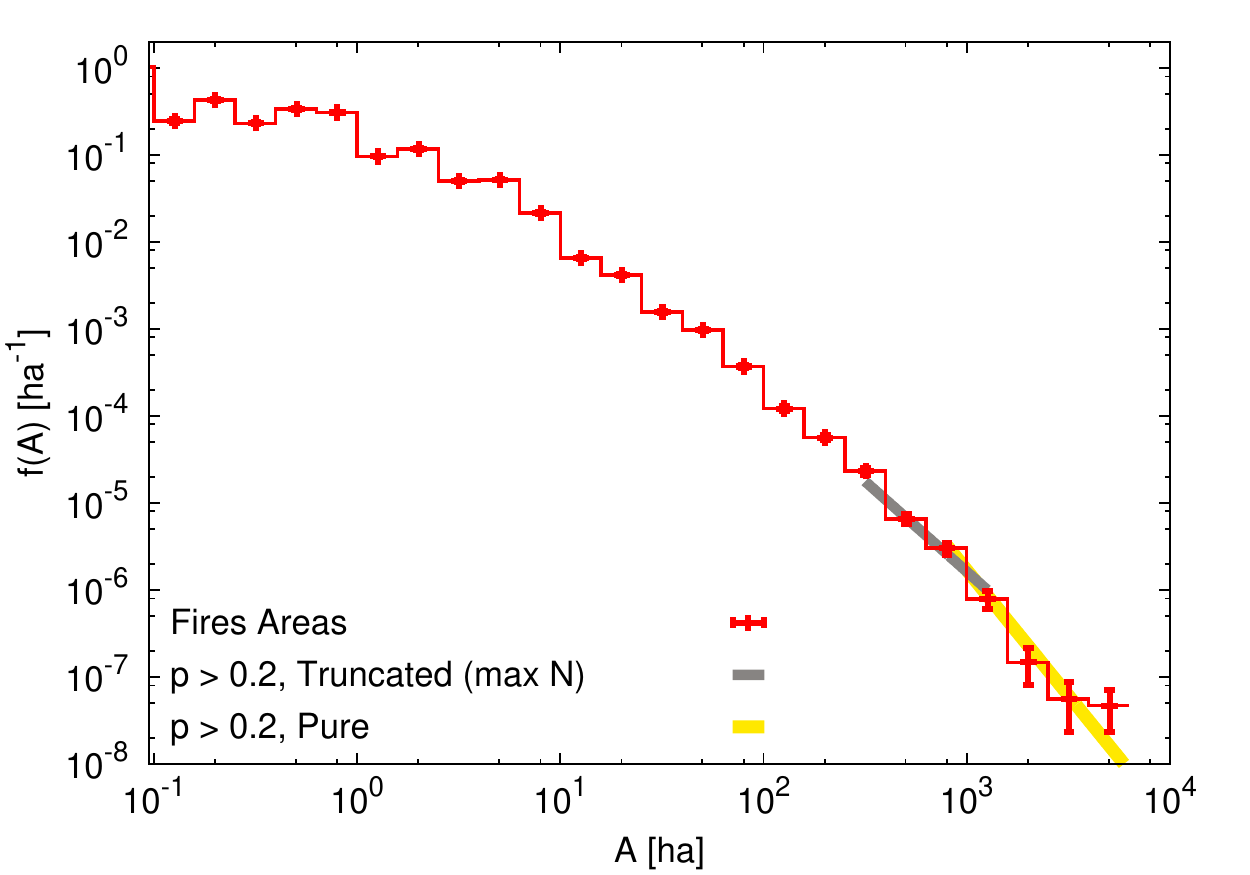}
\caption{Estimated probability density of the area of 
fires in the Italian catalog, together with the power-law fits.
In contrast to the previous datasets analyzed, we consider
these power-law fits as irrelevant.}  
\end{figure}

\begin{table*}[t]
 \caption{
 Results of the fits for the burned area of the $N_{tot}=$ 36 748 fires 
recorded in the Italian catalog, 
for different $p_c$. The cases of a non-truncated power law and a
truncated power law, maximizing $N$, are shown. 
\label{results4}}
\centering
\begin{tabular}{r r r r c c }
 $N$  & $a$ (ha) & $b$ (ha) & ${b}/{a}$ & $\alpha \pm \sigma$ & $p_c$\\ 
\hline 

      51 &            794&       $\infty$ &          $\infty$ & 
   2.880$\pm$ 0.276    & 0.10\\ 
      51 &            794&       $\infty$ &        $\infty$ & 
  2.880$\pm$ 0.277    & 0.20\\ 
      51 &            794&       $\infty$ &          $\infty$ & 
  2.880$\pm$ 0.277    & 0.50\\ 

\hline
    168 &            316&       7943 &       25 &  2.353$\pm$ 0.117    & 0.10\\ 
    148 &            316&       1259 &       4 &  2.075$\pm$ 0.213    & 0.20\\ 
     51 &            794&       79430 &       100 &  2.870$\pm$ 0.281    & 0.50\\ 
\hline
\end{tabular}
\end{table*}

\subsection{Waiting time between earthquakes}

The temporal properties of earthquakes has been
a subject relatively little studied
(at least in comparison with the size of earthquakes).
It is true that the Omori law has been known for more than 
100 years \cite{Utsu_omori,Utsu_handbook}, 
and that this is a law extremely important in order to assess the hazard
of aftershocks after a big event, but the Omori law looks at time properties
in a very coarse-grained way, as it only provides
the number of events in relatively large time windows.
Thus, no information on the fine time structure of seismicity
is provided, at least directly.

The situation has changed in the last decade, 
since the seminal study of \citet{Bak.2002},
who found a unified scaling law for earthquake waiting-time distributions.
They took Southern California and divided it in different areas, 
and computed the time between consecutive earthquakes for each area.
So, if $t_i^j$ denotes the time of occurrence of the $i-$th earthquake
in area $j$, the corresponding waiting time $\tau_i^j$ is
\begin{equation}
\tau_i^j = t_i^j - t_{i-1}^j.
\end{equation}
The key point is that all the resulting waiting times were added
to the same distribution (and not to a different distribution $j$ for
each area).
Subsequently, the 
unified scaling law was found to be valid for other regions 
of the world \cite{Corral_physA.2004}.
The shape of the resulting probability density corresponds to
a double power law, 
one for small waiting times, associated to a mixture of scales
of waiting times due to the Omori law,
and another for large waiting times, due to spatial heterogeneity
arising from the mixture of different areas with different seismic rates
\cite{Bak.2002,Corral_pre.2003,Corral_physA.2004,Corral_Christensen}.
The first exponent was found to be close to 1,
whereas the second one was about 2.2;
the fits were done by means of the 
nonlinear least-squares Marquardt-Levenberg
algorithm from {\tt gnuplot}, applied to the logarithm
of the log-binned empirical density.
Here we apply our more sophisticated method 
to updated data for Southern California seismicity,
with $x=\tau$.

We use again the relocated Southern California catalog
of \citet{Shearer3}, see also \citet{Shearer2},
but starting in 1984 and ending in 
June 30th, 2011.
This is to avoid some holes in the catalog for the preceding years.
As for earthquake sizes, the occurrence takes place
in a rectangle of coordinates ($122^\circ$W,$30^\circ$N), ($113^\circ$W,$37.5^\circ$N).
This rectangle is divided into equal parts both in the West-East axis
and in the South-North axis, in such a way that we consider a number of subdivisions
of $4 \times 4$, $8 \times 8$, $16 \times 16$, and $32\times 32$.
The waiting times 
for events of magnitude $m\ge 2$
in each of these subdivisions are computed as explained above,
resulting in a number of data between 103 000 and 104 000 in all cases.

For a non-truncated power law, the results are only coherent with 
the previous reported ones (exponent around 2.2) for the intermediate cases, 
i.e., $8\times 8$ and $16 \times 16$, see Table VI and Fig. 9.
The disagreement for the other cases can easily be explained.
For $4\times 4$, the number of resulting subdivisions, 16, 
seems rather small. As mentioned, in \citet{Corral_Christensen} the power-law
tail was explained in terms of a power-law mixture of exponentials;
so, with only 16 regions is possible that the asymptotic behavior
is still not reached.
On the other hand, the effect of the finite duration of the catalog 
is visible in the $32\times 32$ data. Due to the scaling behavior of the distributions
\cite{Corral_pre.2003,Corral_physA.2004}, the possible power-law tail in this case
is displaced to larger waiting times; but the time span of the catalog, 
about $10^{10}$ s, clearly alters this power law, 
which starts to bend at about $10^9$ s.
Thus, we conclude that a power-law exponent of about $\alpha\simeq 2.2$ or 2.3
indeed exists, provided that the number of spatial subdivisions is high enough
and the temporal extension of the catalog is large enough.

When a truncated power-law is fitted, using the method of maximizing the number of data $N$,
the other power law emerges, but for a range shorter than what the plot of the densities suggests.
The exponent is in a range from 0.95 to 1.04 (except for the $4 \times 4$ cases, in which it is 
a bit smaller). The largest log-range is 100, i.e., two decades.
The graphical representation of the density seems to indicate that the possible power
law is influenced by the effect of two crossovers, one for large waiting times, 
associated to a change in exponent, and another one for smaller times, 
where the distribution becomes flat.
Finally, the method of fitting which maximizes the log-range
leads to results that are similar to the non-truncated power-law case,
although sometimes intervals corresponding to very small times are selected.
The latter results have no physical meaning, as correspond to times below 1 s,
i.e., below the error in the determination of the occurrence time.

\begin{table*}[t]
 \caption{ 
Results of the fits with a non-truncated power law and a truncated power law,
maximizing $N$, for earthquake waiting times calculated for different
subdivisions of Southern California. Different minimum $p-$values are shown.
The total number of data is above 103 000 in any case.
}
\centering
\begin{tabular}{l r c c r c c  }
Subdivisions & $N$  & $a$ (s) & $b$ (s) & ${b}/{a}$ & $\alpha \pm \sigma$ & $p_c$\\ 
\hline 
$16\times 16$ &     542  &    0.3162 $\times 10^8$ &       $\infty$ &      $\infty$      
&  2.324$\pm$ 0.056    & 0.10\\ 
$32\times 32$ &      67  &   0.3162 $\times 10^9$ &       $\infty$ &      $\infty$     
&  4.404$\pm$ 0.405    & 0.10\\ 
$4\times 4$ &     124  &     0.5012 $\times 10^7$ &       $\infty$ &      $\infty$      
&  1.921$\pm$ 0.085    & 0.10\\ 
$8\times 8$ &    1671  &     0.3162 $\times 10^7$ &       $\infty$ &      $\infty$      
     &  2.198$\pm$ 0.031    & 0.10\\ 
\hline
$16\times 16$ &     542  &    0.3162 $\times 10^8$ &       $\infty$ &      $\infty$     
 &  2.324$\pm$ 0.056    & 0.20\\ 
$32\times 32$ &      67  &   0.3162 $\times 10^9$ &       $\infty$ &      $\infty$      
&  4.404$\pm$ 0.403    & 0.20\\ 
$4\times 4$ &     124  &     0.5012 $\times 10^7$ &       $\infty$ &      $\infty$      
&  1.921$\pm$ 0.085    & 0.20\\ 
$8\times 8$ &    1671  &     0.3162 $\times 10^7$ &       $\infty$ &      $\infty$      
&  2.198$\pm$ 0.031    & 0.20\\ 
\hline
$16\times 16$ &      24  &   0.3162 $\times 10^9$ &       $\infty$ &      $\infty$      
&  4.106$\pm$ 0.703    & 0.50\\ 
$32\times 32$ &      67  &    0.3162 $\times 10^9$ &       $\infty$ &      $\infty$      
&  4.404$\pm$ 0.449    & 0.50\\ 
$4 \times 4$ &      77  &     0.7943 $\times 10^7$ &       $\infty$ &      $\infty$      
&  1.856$\pm$ 0.098    & 0.50\\ 
$8\times 8$ &     322  &    0.1259 $\times 10^8$ &       $\infty$ &      $\infty$      
&  2.231$\pm$ 0.070    & 0.50\\ 

\hline


$16\times 16$ &   44178  &        7943 &       0.7943 $\times 10^6$ &       100 &  0.956$\pm$ 0.004    & 0.10\\ 
$32\times 32$ &   43512  &        1259 &       0.1995 $\times 10^6$ &       158 &  1.029$\pm$ 0.003    & 0.10\\ 
$4\times 4$ &   38765  &        1995 &       0.5012 $\times 10^5$ &       25 &  0.867$\pm$ 0.006    & 0.10\\ 
$8\times 8$ &   39851  &         316 &       0.1995 $\times 10^5$ &       63 &  0.987$\pm$ 0.004    & 0.10\\ 
\hline
$16\times 16$ &   39481  &        7943 &       0.5012 $\times 10^6$ &       63 &  0.950$\pm$ 0.005    & 0.20\\ 
$32\times 32$ &   39654  &        1259 &       0.1259 $\times 10^6$ &       100 &  1.033$\pm$ 0.004    & 0.20\\ 
$4\times 4$ &   38765  &        1995 &       0.5012 $\times 10^5$ &       25 &  0.867$\pm$ 0.006    & 0.20\\ 
$8\times 8$ &   39851  &         316 &       0.1995 $\times 10^5$ &       63 &  0.987$\pm$ 0.004    & 0.20\\ 
\hline
$16\times 16$ &   39481  &        7943 &       0.5012 $\times 10^6$ &       63 &  0.950$\pm$ 0.005    & 0.50\\ 
$32\times 32$ &   39654  &        1259 &       0.1259 $\times 10^6$ &       100 &  1.033$\pm$ 0.004    & 0.50\\ 
$4\times 4$ &   34113  &        3162 &       0.5012 $\times 10^5$ &       16 &  0.864$\pm$ 0.007    & 0.50\\ 
$8\times 8$ &   39851  &         316 &       0.1995 $\times 10^5$ &       63 &  0.987$\pm$ 0.004    & 0.50\\ 
\hline
\end{tabular}
\end{table*}

\begin{figure}[h]
\includegraphics*[width=15cm]{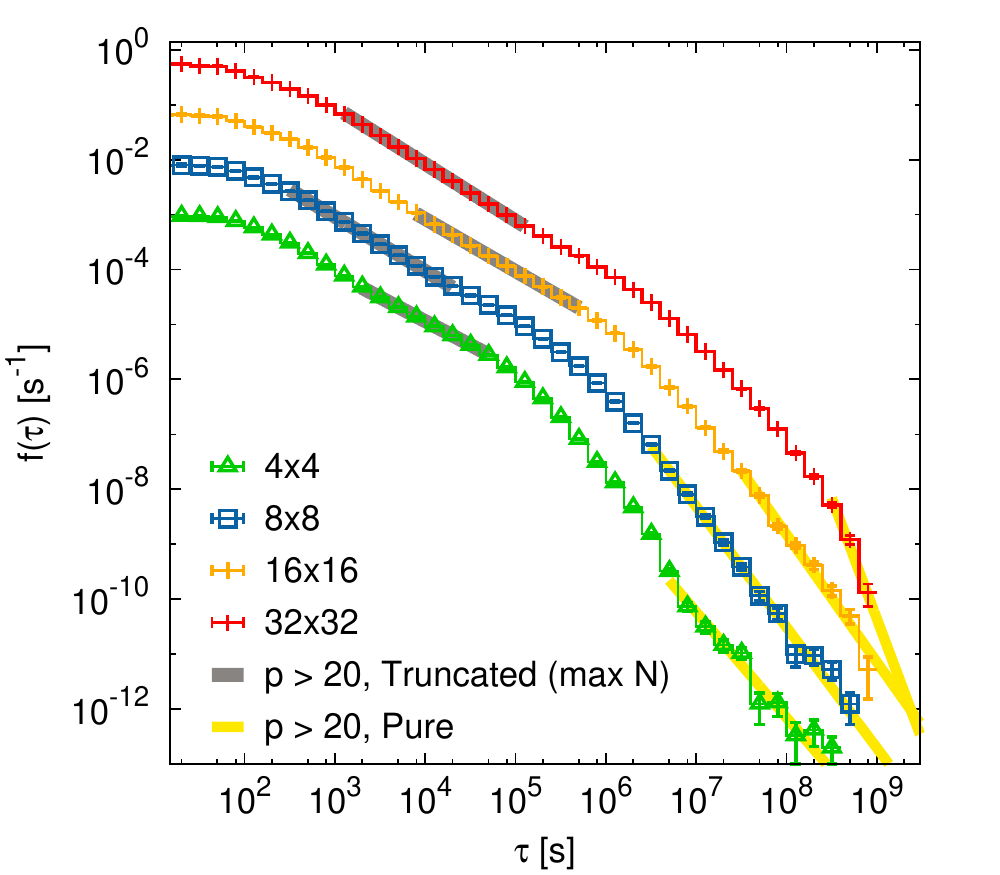}
\caption{
Estimated probability densities and corresponding power-law fits for the waiting times
of $m\ge 2$ in the Southern California catalog, for different spatial subdivisions.
The values of the density are multiplied by factors $1$, $10$, $100$, and $1000$,
for clarity sake.
}  
\end{figure}

\section{Conclusions}

For power-law distributions,
the fitting and testing the goodness-of-fit 
is a difficult but very relevant problem
in complex-systems science, in general,
and in geoscience in particular.
The most critical step is to select,
automatically (without introducing any subjective bias),
where the power-law regime starts and where it ends.
We have explained in detail a conceptually simple but somewhat
laborious procedure in order to overcome some difficulties
previously found in the method introduced by Clauset et al. (2009),
see \citet{Corral_nuclear}.
Our method is summarized in fitting by maximum likelihood estimation
and testing the goodness of fit by the Kolmogorov-Smirnov statistic, 
using Monte Carlo simulations.
Although these steps are in common with the Clauset et al.'s (2009) recipe,
the key difference is in the criterion of selection 
of the power-law range.
Despite the many steps of these procedures,
ours can be easily implemented, 
and the resulting algorithms run very fast
in current computers.
We also have explained how to estimate properly
the probability density of a random variable 
which has a power law or a fat-tail distribution.
This is important to draw graphical representations
of the results of the fitting (specially in Fig. 5)
but it is not required to perform neither the fits
nor the goodness-of-fit tests.

The performance of the method is quite good, 
as checked in synthetic power-law datasets, 
and the results of the analysis of previously reported
power laws are very consistent.
We confirm a very broad power-law tail in the distribution 
of the half-lives of the radionuclides, 
with exponent $\alpha=1.09$,
as well as other power-law regimes in the body
of the distribution.
The results for the power-law exponent 
of the distribution of seismic moments
worldwide and in Southern California are in agreement 
with previous estimates, but in addition
our method provides a reliable way to determining
the minimum seismic-moment value for which
the Gutenberg-Richter law holds.
This can be useful to check systematically
for the completeness thresholds of seismic catalogs.
For the energy dissipated by tropical cyclones,
measured roughly through the PDI, we confirm the power-law behavior 
in the body of the distribution previously reported,
with exponents close to one. We also survey new results
for the Southern Indian Ocean, but with a higher power-law exponent.
In contrast, for the case of the area affected by forest fires 
in an Italian catalog, we obtain power-law-distributed behavior
only for rather small windows of the burnt area, containing few number of data.
Finally, for the waiting times between earthquakes in different
subdivisions of Southern California we conclude that the power-law
behavior of the tail is very delicate, 
affected either by a small number of subdivisions,
when the size of those is large, 
or by the finite duration of the record, 
which introduces a sharp decay of the distribution
when their number is high.
For the body of the distribution another power law
is found, but the range is limited by crossovers
below and above it.
Naturally, our method can be directly applied 
to the overwhelming number of fat-tailed distributions
reported during the last decades in geoscience.

\section*{Acknowledgements}

Our development of the fitting and goodness-of-fit testing method presented 
here was originated from a collaboration with O. Peters, and was influenced
by comments from L. A. N. Amaral and R. D. Malmgren.
Research in hurricanes and half-lives started jointly with A. Oss\'o
and F. Font, respectively.
We got the references for the law of large numbers
from J. Ll. Sol\'e.
L. Telesca and R. Lasaponara provided permission to use
their fire catalog.
We are particularly indebted to I. Serra and J. del Castillo,
and to Per Bak, who is the ultimate ancestor of all this.
Research projects providing some money to us were
FIS2009-09508, from the disappeared MICINN,
FIS2012-31324, from MINECO,
and 2009SGR-164,
from Generalitat de Catalunya.


\bibliography{biblio}

\end{document}